\documentclass{aa}

\usepackage{graphicx}
\usepackage{amsmath,amsfonts}

\usepackage{natbib} %option sort&compress pour respecter l'ordre du fichier .bib
\bibpunct{(}{)}{;}{a}{}{,} % A&A bibliography style

\newcommand{\modif}[2][modif]{{#2}} % modifications invisibles
\begin{document}
  
%%%\thesaurus{05(03.13.4; 05.03.1; 11.14.1; 11.19.4)}

  \title{A new Monte Carlo code for star cluster simulations: I.~Relaxation}
  %\subtitle{}

  \author{M. Freitag\inst{1}\thanks{\emph{Present address:}\ Physikalisches Institut, 
  Universit\"at Bern, 
  Sidlerstrasse 5,
  CH-3012 Bern,
  Switzerland} 
  \and W. Benz\inst{2}}
  
\institute{
  Observatoire de Gen\`eve,
  Chemin des Maillettes 51,  
  CH-1290 Sauverny,
  Switzerland
  \and Physikalisches Institut, 
  Universit\"at Bern, 
  Sidlerstrasse 5,
  CH-3012 Bern, 
  Switzerland}
  
\offprints{M. Freitag}
\mail{Marc.Freitag@obs.unige.ch}
  
\date{Received 8 February 2001/ Accepted 23 April 2001}
  
\titlerunning{Monte Carlo Cluster Simulations~I}
\authorrunning{M. Freitag \& W. Benz}

\abstract{ We have developed a new simulation code aimed at studying the
  stellar dynamics of a galactic central star cluster surrounding a
  massive black hole. In order to include all the relevant physical
  ingredients (2-body relaxation, stellar mass spectrum, collisions,
  tidal disruption, \ldots), we chose to revive a numerical scheme
  pioneered by H\'enon in the 70's \citep{Henon71a,Henon71b,Henon73}.
  It is basically a Monte Carlo resolution of the Fokker-Planck
  equation. It can cope with any stellar mass spectrum or velocity
  distribution.  Being a particle-based method, it also allows one to take
  stellar collisions into account in a very realistic way. This first
  paper covers the basic version of our code which treats the
  relaxation-driven evolution of a stellar cluster without a central BH.
  A technical description of the code is presented, as well as the
  results of test computations. Thanks to the use of a binary tree to
  store potential and rank information and of variable time steps,
  cluster models with up to $2\times10^6$ particles can be simulated
  on a standard personal computer and the CPU time required scales
  as $N_\mathrm{p}\ln(N_\mathrm{p})$ with the particle number
  $N_\mathrm{p}$. Furthermore, the number of simulated stars needs not
  be equal to $N_\mathrm{p}$ but can be arbitrarily larger. A
  companion paper will treat further physical elements, mostly
  relevant to galactic nuclei.
  \keywords{methods: numerical -- stellar dynamics --
    galaxies: nuclei -- galaxies: star clusters} }
  
  \maketitle

\section{Introduction} 
\label{sec:intro}

The study of the dynamics of galactic nuclei is likely to become a
field to which more and more interest will be dedicated in the near
future. Numerous observational studies, using high resolution
photometric and spectroscopic data, point to the presence of
``massive dark objects'' in the centre of most bright galaxies
\citep[see reviews
by][]{KR95,Rees97,FTFJ98,Mag98,Ho99,VdM99a,deZeeuw00}, including our
own Milky Way \citep{EG96,EG97,GEOE97,GKMB98,GMB99a,GMB99b,GPEGO00}.
As the precision of the measurements and the rigor of their
statistical analysis both increase, the conclusion that these central
mass concentrations are in fact black holes (BHs) with masses ranging
from $10^6\,M_{\sun}$ to a few $10^9\,M_{\sun}$ becomes difficult to
evade, at least in the most conspicuous cases \citep{Maoz98}. Hence,
the time is ripe for examining in more detail the mutual influence
between the central BH and the stellar nucleus in which it is
engulfed.

A few particular questions we wish to answer are the following:

\begin{itemize}
\item What is the stellar density and velocity profile near the
  centre?
\item What are the stellar collision and tidal disruption rates? 
  Were they numerous enough in past epochs to contribute
  significantly to the growth of the central BH? Do they occur
  frequently enough at present to be detected in large surveys
  and thus reveal the presence of the BH?
\item What dynamical role do these processes play? Are they just
  by-products of the cluster's evolution or are there any feed-back
  mechanisms?
\item What is the long-term evolution of such a stellar system? Will
  it experience core collapse, re-expansion, or even gravothermal 
  oscillations like globular clusters?
\end{itemize}

To summarize, our central goal is to simulate the evolution of a
galactic nucleus over a few $10^9$\,years while taking into account a
variety of physical processes that are thought to contribute
significantly to this evolution or are deemed to be of interest for
themselves. Here is a list of the ingredients we want to incorporate
into our simulations:

\begin{itemize}
\item Relaxation induced by 2-body gravitational encounters. The
  accumulation of these weak encounters leads to a redistribution of
  stellar orbital energy and angular momentum. The role of relaxation
  has been extensively studied in the realm of globular clusters where
  it leads to the slow built-up of a diffuse halo and an increase in
  the central density until the so-called ``gravo-thermal''
  catastrophe sets in (see Sect.~\ref{subsec:plummer_cc}). Of
  particular relevance to galactic nuclei are early studies that
  demonstrated that, when a central star-destroying BH is present,
  relaxation will lead to the formation of a quasi-stationary cusp in
  the stellar density
  \citep{Peebles72,SL76,BW76,DO77a,DO77b,LS77,SM78,CK78}. 
\item Tidal disruptions of stars by the BH \citep{Rees88,Rees90}. Not
  only can this be a significant source of fuel for the central BH,
  but it can also play a role in the cluster's evolution, as specific
  orbits are systematically depleted. The destruction of deeply bound
  stars may lead to heating the stellar cluster \citep{MS80}.
\item Stellar collisions. As the stellar density increases in the
  central regions, collisions become more and more frequent.
  \modif[3]{By comparing the relaxation time and the collision time
    \citep[ for instance]{BT87}, one sees that collisions can catch up
    with relaxation when $\sigma_v >
    \left(\ln\Lambda\right)^{1/4}V_{\ast} \approx
    1000\,\mathrm{km}\,\mathrm{s}^{-1}$, where $\sigma_v$ is the
    velocity dispersion in the stellar cluster, $\ln\Lambda$ the
    Coulomb logarithm and $V_{\ast}$ the escape velocity from the
    surface of a typical main sequence (MS) star.} But even in colder
  clusters, collisions may be of interest as a channel to create
  peculiar stars (blue stragglers, stripped giants,\ldots).
\item Stellar evolution. As stars change their masses and radii, the
  way they are affected by relaxation, tidal disruptions and
  collisions is strongly modified. Of particular importance is the
  huge increase in cross section during the giant phase and its nearly
  vanishing when the star finally evolves to a compact remnant. As
  they are not prone to being disrupted, these compact stars may
  segregate towards the centre-most part of the nucleus where they may
  dominate the density \citep[ amongst others]{Lee95,MEG00}.
  Furthermore, the evolutionary mass loss (winds, planetary nebulae,
  SNe\ldots) may strongly dominate the BH's growth if this gas sinks
  to the centre \citep{DDC87b,NS88,MCD91}.
\item BH growth.  Any increase of the BH's mass may obviously lead to
  a higher rate of tidal disruptions. It also imposes higher stellar
  velocities and, thus, increases the relative importance of collisions
  in comparison with relaxation. A further contributing effect is
  the central density built-up due to the adiabatic adaptation of
  stellar orbits to the deepening potential
  \citep{Young80,LG89,CB94,QHS95,SHQ95,LA00}.
\end{itemize}

Although we restrict our discussion to these physical ingredients in
the first phase of our work, there are many others of potential
interest; some of them are mentioned in Sect.~\ref{subsec:future}.

Thus, the evolution of galactic nuclei is a complex problem. As it is
mostly of stellar dynamical nature, our approach is grounded on a new
computer code for cluster dynamics. Its ``core'' version treats the
relaxational evolution of spherical clusters. This is the object of the
present paper. The inclusion of further physical ingredients such as
stellar collisions and tidal disruptions will be explained in a
complementary article (\citealt{FB01b}, see also
\citealt{FB01d,FB01e}). In order to obtain a realistic prescription for
the outcome of stellar collisions, we compiled a huge database of
results from collision simulations performed with a Smoothed Particle
Hydrodynamics (SPH) code \citep{Benz90}. This work will be described
in a third paper \citep{FB01c}.

This paper is organized as follows. In the next section, we briefly
review the available numerical schemes that have been applied in the
past to simulate the evolution of a star cluster and explain the
reasons that led us to choose one of these methods.
Sections~\ref{sec:general} to \ref{sec:other} contain a detailed
description of the code. Test calculations are presented in
section~\ref{sec:tests}.  Finally, in section~\ref{sec:concl}, we
summarize our work and discuss future improvements.

\section{Choice of a simulation method}
\label{sec:sim_methods}

\modif[4]{In the previous section, we briefly reviewed the main
  processes that should play an important role in the evolution of
  galactic nuclei. To treat them realistically, any stellar dynamics
  code has to meet the following specifications. It must:}

\begin{itemize}
\item Allow for non-isotropic velocity distributions. This is
  especially important to realistically describe the way tidal
  disruptions deplete very elongated orbits (the loss-cone
  phenomenon).
\item Incorporate stars with different masses. Otherwise, mass
  segregation would be neglected and considering the outcome of
  stellar collisions properly would be impossible. Furthermore, it
  must be possible to make the stellar properties (mass, radius)
  change with time to reflect stellar evolution.
\item Be able to evolve the cluster over several relaxation
  time-scales. The amount of required computing (``CPU'') time can be
  kept to a reasonable level only if individual time steps are a
  fraction of the relaxation time scale rather than the orbital
  period.
\end{itemize}

Several techniques suitable for simulations of cluster evolution have
been proposed in the literature.  Many textbooks and papers review
these methods so we can restrict ourselves to a short overview
\citep{BT87,Spitzer87,MH97,HGST99,Spurzem99,SK99}.

As they directly integrate the particles' equations of motion,
$N$-body simulations are conceptually straightforward and do not rely
on any important simplifying assumptions. Unfortunately, to correctly
compute relaxation effects, forces have to be evaluated by direct
summation \citep[see, however,][ for possible
alternatives]{JP89,McMA93}. Hence, even with specialized hardware such
as GRAPE boards \citep{Makino96,TMFES96,SA96}, following the
trajectory of several $10^6$ stars over tens of relaxation times
remains impossible \citep[but see][]{Makino00} for a peek at the
future of such systems). A solution would be to extrapolate the
results of an $N$-body simulation with a limited number of particles
to a higher $N$. However, this has been shown to be both tricky and
risky \modif[5]{\citep{McMillan93,Heggie97b,AH98,HGST99}}. The difficulty resides
in the fact that various processes (relaxation, evaporation, stellar
evolution\ldots) have time scales and relative importances that depend
in different ways on the number of simulated stars. Thus, in
principle, simulating a cluster of $N_\ast$ stars with a lower number
of particles, $N\ll N_\ast$, could only be done if all these processes
are somehow scaled to the correct $N_\ast$. Unfortunately, such
scalings are difficult to apply, precisely because the $N$-body method
treats gravitation in a direct, ``microscopic'' way with very little
room for arbitrary adjustments.  Furthermore, these modifications
would rely on the same kind of theory, and hence, of simplifying
assumptions, as other simulation methods.

To circumvent these difficulties, a class of methods has been
developed in which a stellar cluster containing a very large number of
bodies is regarded as a fluid \citep[ amongst many
others]{Larson70a,Larson70b,LBE80,LS91,GS94,GS00}. Such simulations
have proved to be very successful in discovering new phenomena like
gravo-thermal oscillations \citep{BS84} but rely on many strong
simplifying assumptions. Most of them are shared by Fokker-Planck and
Monte Carlo methods (see next paragraphs). However, this approach,
which is based on velocity-moments of the collisional Boltzmann
equation, further assumes, as a closure relation, some local
prescription for heat conduction through the stellar cluster. This is
appropriate for a standard gas, where the collision mean free path
is much smaller than the system's size \citep{HSBT95}. Quite
surprisingly, it still seems to be valid for a stellar cluster even
though the radial excursion of a typical star is not ``microscopic''
\citep{BS85,GS94,ST95}. However, we discarded such an approach
because, due to its continuous nature, integrating the effects of
collisions in it seems difficult.

A very popular intermediate approach is the ``direct'' numerical
resolution of the Fokker-Planck equation (FPE) \citep[ for
example]{BT87,Spitzer87,Saslaw85}, either by a finite difference
scheme \citep[ and many other
works]{Cohn79,Cohn80,Lee87,MCD91,Takahashi95,DCLY99} or using finite
elements \citep{Takahashi93,Takahashi95}. Here again, the main
difficulty resides in the treatment of collisions. From a practical
point of view, these methods represent the cluster as a small set of
continuous distribution functions for discrete values of the stellar
mass. A realistic modeling of collisional effects would then require one
to multiply the number of these mass classes, at the price of an
important increase in code complexity and computation time. From a
theoretical point of view, collisions don't comply with the
requirement of small orbital changes that is needed to derive the FPE.

So, we have finally turned our attention to the so-called ``Monte
Carlo'' (MC) schemes. Even though their underlying hypotheses are
similar to those leading to the FPE, being particle methods, they
inherit some of the $N$-body philosophy which allows us to extend their
realm beyond the set of problems to which direct FPE resolutions apply.
In the Monte Carlo approach, the evolution of the (spherically
symmetric) cluster is computed by following a sample of test-particles
that represent spherical star shells. They move according to the
smooth overall cluster (+BH) gravitational potential and are given
small orbital perturbations in order to simulate relaxation. These
encounters are randomly generated with a probability distribution
chosen in such a way that they comply with the diffusion coefficients
appearing in the FPE.

The most recently invented MC code is the ``Cornell'' program by
Shapiro and collaborators \citep[ and references therein]{Shapiro85}
with which these authors conducted seminal studies of the evolution a
star cluster hosting a central BH. At the end of each time step, the
distribution function (DF) is identified with the distribution of
test-particles in phase-space.  Then the potential is re-computed and
so are diffusion coefficients that are tabulated over phase-space.
This allows us to evolve the DF one step further by applying a new series
of perturbations to the test-particles' orbits. This code features
ingenious improvements, such as the ability to follow the orbital motion
of test-particles threatened by tidal disruption and to ``clone''
test-particles in the central regions in order to improve the
numerical resolution. Despite its demonstrated power, we did not adopt
this technique because of its already important complexity, which would
increase significantly if additional physics such as stellar
collisions, a stellar mass spectrum, etc, are introduced.

Spitzer and collaborators \citep{Spitzer75,SS75a,SS75b,SM80} and
H\'enon \citeyearpar{Henon66,Henon71a,Henon71b,Henon73} developed
simpler MC schemes which show more potential for our purpose. The
simulation of relaxation proceeds through 2-body gravitational
encounters between neighboring shell-like particles. The deflections
are tailored to lead to the same diffusion of orbits that a great
number of very small interactions would cause in a real cluster.
After the encounter, the shell is placed at some radius $R$ on its
modified orbit. At that time, the potential is updated so it remains
consistent with the positions of the shells. The main difference
between the ``Princeton'' code by Spitzer~et~al. and H\'enon's
algorithm is that the former uses a fraction of the orbital time as
its time step while the latter uses a fraction of the relaxation time.
It follows that out-of-equilibrium dynamical processes, like violent
relaxation, can be computed with the Princeton code whereas H\'enon's
code can only be applied to systems in virial equilibrium, which
imposes an age greatly in excess of the relaxation time. Needless to
say, this restriction is rewarded by an important gain in speed. In
this scheme, instead of being determined by the equations of orbital
motion, the radius $R$ of a given shell is randomly chosen according
to a probability distribution that measures the time spent at each
radius on its orbit: $\mathrm{d}P/\mathrm{d}R \propto
1/v_{\mathrm{rad}}(R)$, where $v_{\mathrm{rad}}(R)$ is its radial
velocity at $R$. $R$-dependent time steps can be used to track the huge
variation of the relaxation time from the cluster's centre to its
outskirts.

We chose to follow the approach developed by H\'enon to write our own
Monte Carlo code. It indeed appeared as an optimal compromise in
terms of physical realism and computational speed. On the one hand, it
allows for all the key physical ingredients listed at the beginning of
this section. On the other hand, high resolution simulations are
carried out in a few hours to few days on a standard personal
computer. This will enable a wide exploration of the parameter space.

Since H\'enon's work, this approach has been extensively modified and
successfully applied to the study of the dynamical evolution of
globular clusters by Stod\'{o}{\l}kiewicz
\citeyearpar{Stodol82,Stodol86} and Giersz
\citeyearpar{Giersz98,Giersz00,Giersz00b}. Another H\'enon-like MC
code has recently been written by Joshi~et~al.
\citep{JRPZ00,Rasio00,WJR00,JNR01}. As far as we know, however, no one
ever applied this simulation method to galactic nuclei.

The Monte Carlo scheme relies on the central assumption that the
stellar cluster is always in dynamical equilibrium. This is the case
for well relaxed systems. Sufficient observational resolution has only
recently been obtained to allow an estimate of the relaxation times in
the nearest galactic nuclei.  \citet{Lauer98} report relaxation times
$T_\mathrm{rel}$ of about $7\times 10^{11}\,\mathrm{yrs}$, $3\times
10^9\,\mathrm{yrs}$ and $3\times 10^6\,\mathrm{yrs}$ at $0.1\,
\mathrm{pc}$ for M31, M32 and M33 respectively.  As for the Milky Way,
\citet{Alexander99} deduces $T_\mathrm{rel} \simeq 3\times 10^9
\,\mathrm{yrs}$ at $0.4\,\mathrm{pc}$ (core radius), a value that does
not change significantly at smaller radii if a $\rho \propto R^{-1.8}$
cusp model is assumed with the parametrisation of \citet{GTKKTG96} for
the velocity dispersion.  \citet{GHT94} get $T_\mathrm{rel} \simeq
4\times 10^7\,\mathrm{yrs}$ for the central value but the meaning of
this value is unclear, as the dynamical influence of the BH was
neglected in its derivation. These few values indicate that, perhaps,
only a subset of all galactic nuclei are amenable to the kind of
approach we are developing. However, these very dense environments
with relaxation times lower than the Hubble time are precisely the
ones which expectedly lead to non-vanishing rates for the disruptive
events we are primarily interested in.  \modif[7]{Furthermore, note that,
contrary to the case of globular clusters, stellar evolution of a
population of massive stars (e.g. in a starburst) can probably not
disrupt dynamical equilibrium (through mass-loss effects) as its time
scale ($\ge$ a few $10^6$\,yrs) is longer than orbital periods in a
galactic nucleus ($\le$ a few $10^5$\,yrs).}

\section{General considerations}
\label{sec:general}

\subsection{Principles underlying code design}

In our work, we focus at the long term evolution of star clusters, on
time scales much exceeding the dynamical (crossing) time,
$T_\mathrm{dyn} \simeq \sqrt{R_\mathrm{cl}^3/(GM_\mathrm{cl})}$, where
$M_\mathrm{cl}$ is the cluster's total mass and $R_\mathrm{cl}$ a
quantity indicating its size (for instance the half-mass radius). We
thus make no attempt at describing evolutionary processes that occur
on a $T_\mathrm{dyn}$ time scale, most noticeably phase mixing and
violent relaxation, which are thought to rule early life phases of
stellar systems (see, for instance, section 4.7 of \citealt{BT87} or
section 5.5 of \citealt{MH97} and references therein). Hence, we
assume that the cluster has reached a state of dynamical equilibrium.
Its subsequent evolution, driven by processes with time scales $\gg
T_\mathrm{dyn}$ (2-body relaxation, collisions, tidal disruptions and
stellar evolution), passes through a sequence of such states.

This reasonable restriction allowed H\'enon to devise a simulation
scheme whose time step is a fraction of the relaxation time instead of
the dynamical time. Naturally, this leads to an enormous gain in
computation speed compared to codes that resolve orbital processes
like $N$-body programs or the Princeton Monte Carlo code
\citep{SH71a,ST72}.

Another strong simplifying assumption the scheme heavily relies on is
that of spherical symmetry. This makes the cluster's structure
effectively one-dimensional which allows a simple and efficient
representation for the gravitational potential (see
Sect.~\ref{sec:pot_rep}) and the stars' orbits and furthermore leads to
a straightforward determination of neighboring particle pairs. Of
course, such a geometry greatly facilitates the computation of any
quantity describing the cluster's state, such as density, velocity
dispersion and so on. An obvious drawback of this assumption is that
it forbids the proper treatment of any non-spherical feature as
overall rotation \citep{Arab97,Einsel98,ES99,SE99,Boily00}, an
oscillating or a binary black hole
\citep{BBR80,LT80,Makino97,QH97,GR00,MCM00} or an accretion disk
interacting with the star cluster
\citep{SCR91,Rauch95,AZD96,VK98a,VK98b}.

In H\'enon's scheme, the numerical realization of the cluster is a set
of spherical thin shells, each of which is given a mass $M$, a radius
$R$, a specific angular momentum $J$ and a specific kinetic energy
$T$. \modif[8]{In the remainder of the paper, we refer to these %ICI
  particles as ``super-stars'' because we regard them as spherical
  layers of synchronized stars that share the same stellar properties,
  orbital parameters and orbital phase, whose mass is spread over the
  shell and which experience the same processes (relaxation,
  collision, \ldots) at the same time. In recent implementations of
  H\'enon's method by \citet{Giersz98} and \citet{JRPZ00}, each
  particle represents only one star. This avoids scaling problems
  connected with the computation of the rate of 2- (or many) body
  processes but would impose too large a number of particles for
  galactic nuclei simulations ($10^6-10^8$ stars). In our code, each
  super-star consists of many stars.  Hence, a cluster containing
  $N_\ast$ stars may be represented by a smaller number of
  super-stars, $N_{\mathrm{SS}} < N_\ast$. The number of stars per
  super-star, $N_\ast/N_{\mathrm{SS}}$, is the same for every
  super-star, in order to conserve energy and angular momentum (as
  well as mass when collisions are included) when simulating 2-body
  processes.}

The super-stars' radii being known, the potential can be computed at
any time and any place so that the orbital energies of all super-stars
are straightforwardly deduced from their kinetic energies and
positions. Hence the set of super-stars can be regarded as a
discretized representation of the one-particle distribution function
(DF) $f$ which, by virtue of Jean's theorem, only depends on the specific
orbital energy $E$ and angular momentum $J$:
\[
    f(\vec{x},\vec{v}) = F(E(\vec{x},\vec{v}),J(\vec{x},\vec{v}))
    = F(\Phi_\mathrm{s}(R)+\frac{1}{2}v^2,Rv_{\mathrm{tg}})
\]
where $\vec{x}$ and $\vec{v}$ are the position and velocity of a star,
$R=\|\vec{x}\|$, $v=\|\vec{v}\|$ and $v_{\mathrm{tg}}$ is the
tangential component of $\vec{v}$. However, whereas a functional
expression of the DF, although a complete description of the stellar
system\footnote{More precisely, the DF, being continuous, is an
  idealized (or statistical) description of the $N_\ast$ star
  system.}, would impose lengthy integrations (resolution of Poisson
equation, as needed in direct FP methods) to yield the gravitational
potential, the Monte Carlo representation of the cluster provides it
directly. From this point of view, the Monte Carlo method is closer to
an $N$-body philosophy than to direct FP methods.

The main difference between the MC code and a spherical 1D $N$-body
simulation is that the former does not explicitly follow the
continuous orbital motion of particles which preserves $E$ and $J$.
However these orbital constants, as well as other properties of the
super-stars, are modified by ``collisional'' processes to be
incorporated explicitly, namely 2-body relaxation, stellar collisions
and tidal disruptions. So, in the version of the code described here,
the MC simulation proceeds through millions to billions of steps, each
of them consisting of the selection of super-stars
(Sect.~\ref{sec:Rel_Pair_Choice}), the modification of their properties
to simulate the effects of relaxation (Sect.~\ref{sec:MC_relax_sim})
and the selection of radial positions corresponding to their new
orbits (Sect.~\ref{sec:selecRorb}).

\subsection{Physical units}
\label{sec:units}

In the rest of this paper, unless otherwise stated, we use the
``$N$-body'' units as prescribed by \citet{HM86}. In this system,
\[
\begin{array}{lcll}
  G &=&1&\mbox{\ \ (Gravitational constant)},\\
  M_0 &=&1&\mbox{\ \ (Total initial cluster mass)},\\
  E_0 &=&-1/4&\mbox{\ \ (Total initial cluster energy)}.
\end{array}
\]
It follows that the corresponding unit of length,
$\mathcal{U}_\mathrm{l}$ and unit of time,
$\mathcal{U}_\mathrm{t}$, can be expressed in any system
of units as
\begin{equation}
  \mathcal{U}_\mathrm{l} = \frac{GM_0^2}{-4E_0} \mbox{\ \
    and\ \ } \mathcal{U}_\mathrm{t} =
  \frac{GM_0^{5/2}}{\left(-4E_0\right)^{3/2}}.
\end{equation}
If the initial cluster's total gravitational energy is written as
\begin{equation}
  U_0 = -\frac{q}{2} \frac{GM_0^2}{R_\mathrm{h}},
\end{equation}
where $q$ is a dimension-less constant and $R_\mathrm{h}$ is the
half-mass radius, we get, for the time unit,
\begin{equation}
  \mathcal{U}_\mathrm{t} = q^{-3/2} 
  \sqrt{\frac{R_\mathrm{h}^3}{GM_0}}.
\end{equation}

For instance, for the often used Plummer model, with stellar density
$\rho(R) \propto \left(1+(R/R_{\mathrm{P}})^2\right)^{-5/2}$, the ``$N$-body'' units
are:
\[
  \mathcal{U}_\mathrm{l} =
  \left(\frac{16}{3\pi}\right)R_\mathrm{P} 
  \mbox{\ \ and\ \ } 
  \mathcal{U}_\mathrm{t} =
    \left(\frac{16}{3\pi}\right)^{3/2} \!\! \sqrt{\frac{R_\mathrm{P}^3}{GM_0}}. 
\]
As we study systems that are stationary on orbital time scales and
whose long-term evolution is driven by relaxation, it is more adequate
to adopt a time unit that scales with relaxation time rather than
dynamical time:
\begin{equation}
  \tilde{\mathcal{U}}_\mathrm{t} = \frac {N_\ast} {\ln(\gamma
    N_\ast)} \mathcal{U}_\mathrm{t}= 7.25\,q^{-3/2}
  T_\mathrm{rel}^\mathrm{h}
  \label{eq:time_unit_rel}
\end{equation}
where $T_\mathrm{rel}^\mathrm{h}$ is the standard ``half-mass'' relaxation
time \citep[ p.~40]{Spitzer87}.

See next section for further explanations of the
relaxation time and, in particular the value of $\gamma$.

\section{Relaxation.}
\label{sec:relax}

\subsection{Summary of standard relaxation theory.}
\label{sec:std_relax_theory}

The theory of relaxation in stellar systems can be found in classical
textbooks \citep[ for instance]{Henon73,Saslaw85,Spitzer87,BT87} and
will not be presented here.  However, the treatment of relaxation is
the backbone of H\'enon's Monte Carlo scheme. Hence a short summary of
these issues is particularly worthwhile, not only to expose the inner
workings of the MC method, but also to understand the limitations it
suffers from (as other statistical cluster dynamics approaches) that
stem from relaxation theory's simplifying assumptions.  Furthermore,
its specific advantages are also to be explained in that framework.

\modif[10]{The basic idea behind the concept of relaxation is that the
gravitational potential of a stellar system containing a large number
of bodies can be described as the sum of a dominating smooth
contribution ($\Phi_\mathrm{s}$) plus a small ``granular'' part
($\delta\Phi$). When only the former is taken into account, the
phase-space DF of the cluster
obeys the collisionless Boltzmann equation.  In the long run,
however, the fluctuating $\delta\Phi$ makes $E$ and $J$ slowly change
and the DF evolve. The basic simplifying assumption underlying
relaxation theory is to treat the effects of $\delta\Phi$ as the sum
of multiple uncorrelated 2-body hyperbolic gravitational encounters
with small deviation angles. Under these assumptions, if a test star
``1'' travels through a homogeneous field of stars ``2'' which all
share the same properties (masses and velocities) during $\delta t$,
its trajectory will deviate from the initial direction by an angle
$\theta$ with the following statistical properties:
\begin{eqnarray}
  \left\langle \theta \right\rangle_{\delta t} &=&0, \nonumber \\
  \left\langle\theta^2\right\rangle_{\delta t} &\simeq& 
  8\pi n \ln\left(\frac{b_\mathrm{max}}{b_0} \right) 
  \frac{G^2\left(M_1+M_2\right)^2}{v_\mathrm{rel}^3} \: \delta t. \label{eq:theta_2_code}
\end{eqnarray}

In this equation, $n$ is the number density of stars, $M_1$ the mass
of the test star, $M_2$ the mass of each field-star, $v_\mathrm{rel}$
the relative velocity between the test star and the field stars, $b_0$
is the impact parameter leading to a deviation angle of $\pi/2$ and
$b_\mathrm{max}$ is a cut-off parameter needed to avoid a logarithmic
divergence. This ill-defined value represents the largest possible
impact parameter and is thus expected to be of the order of the size
of the stellar system, $R_\mathrm{cl}$. If $\sigma_v$ is the velocity
dispersion in the cluster and $M_\ast$ the average stellar mass, the
argument of the so-called ``Coulomb logarithm'' can be approximated by
\begin{equation}
  \frac{b_\mathrm{max}}{b_0} \simeq
  \frac{v_\mathrm{rel}^2 R_\mathrm{cl}}{G\left(M_1+M_2\right)}
  \simeq \frac{\sigma_v^2 R_\mathrm{cl}}{G M_\ast} \simeq \gamma \frac{M_\mathrm{cl}}{M_\ast}=\gamma N_\ast
  \label{eq:log_coul}
\end{equation}
where $\gamma$ is a non-dimensional proportionality constant. This
proportionality only holds for a self-gravitating, virialized
cluster. The exact value of $b_\mathrm{max}/b_0$ is of little
importance as it is embedded into a logarithm. Hence, in most
applications, a constant $\gamma$ is used whose value is determined
either from theoretical arguments \citep{SH71a,Henon75} or by $N$-body
simulations \citep{GH94a,GH96,DCLY99}. The latter approach supports
the classical ``weak encounters'' relaxation theory described here by
showing good agreement with it for properly fitted $\gamma$ values.
Furthermore it confirms \citet{Henon75} who derived $\gamma \simeq
0.10 - 0.17$ for single mass clusters and demonstrated the need for a
much smaller value when an extended stellar mass spectrum is treated.
Here, we use $\gamma = 0.14$ and $\gamma = 0.01$ respectively.

As Eq.~\ref{eq:theta_2_code} is of central importance for the
simulation of relaxation in the Monte Carlo code, we comment on the
main assumptions on which its derivation relies. First, as already
stated, deflections are assumed to be of very small amplitude and
uncorrelated with each other.  The contribution of encounters with
impact parameters between $b_1$ and $b_2$ is of order
$\ln\left(b_2/b_1\right)$ so equal logarithmic intervals of $b$
contribute equally to $\left\langle\theta^2\right\rangle$ and most of
the relaxation is indeed created by ``distant encounters'': $b \gg b_0
\Rightarrow \theta \ll \pi/2$.}

Moreover, the derivation applies in principle only to homogeneous
systems with a finite size. However, a real cluster is grossly
inhomogeneous, with large density gradients.  Applying
Eq.~\ref{eq:theta_2_code} for realistic clusters forces us into
assuming the ``local approximation'', i.e.  stating that typically $b
\ll R_\mathrm{cl}$.  Then, not only can we neglect the effect of
$\Phi_\mathrm{s}$ during an encounter and treat the trajectories as
Keplerian hyperbolae, but, as an added benefit, we can use the
\emph{local} properties of the cluster (density and velocity
distribution) as representative of field stars met by the
test-particle. Admittedly, this is a bold assumption only partially
justified by the ``$\ln(b_2/b_1)$'' argument. The validity of these
approximations has been assessed by comparing results of codes based
on relaxation theory with $N$-body simulations which do not rely on
such assumptions
\citep{GH94a,SA96,Giersz98,PZHMMcM98,TPZ98,Spurzem99}.

Recasting Eqs.~\ref{eq:theta_2_code}, \ref{eq:log_coul} into
\begin{equation}
  \left\langle\theta^2\right\rangle_{\delta t} \simeq
  \left(\frac{\pi}{2}\right)^2 \frac{\delta t}
  {\widehat{T}_\mathrm{rel}^{(1,2)}},
  \label{eq:theta_2_code2}
\end{equation}
we define
\begin{equation}
  \widehat{T}_\mathrm{rel}^{(1,2)} = \frac{\pi}{32} \frac{v_\mathrm{rel}^3}
  {\ln\left(\gamma N_\ast\right) G^2 n \left(M_1+M_2\right)^2}.
  \label{eq:trel_encounter}
\end{equation}
We call this quantity the ``encounter relaxation time'' to insist on
its depending on the properties of one particular class of encounters,
namely those between stars of mass $M_1$ and stars of mass $M_2$ of
(local) density $n$ with relative velocity $v_\mathrm{rel}$. It can be
loosely interpreted as the time needed for encounters with stars ``2''
to gradually deflect the direction of motion of star ``1'' by a RMS
angle $\pi/2$.

\subsection{Monte Carlo simulation of relaxation.}
\label{sec:MC_relax_sim}

\subsubsection{Elementary numerical encounter.}
\label{sec:super_encounter}

Contrary to Fokker-Planck codes, H\'enon's method was devised to avoid
the computational burden and the necessary simplifications connected
with the numerical evaluation of diffusion coefficients (DCs). It does
so through a direct use of Eq.~\ref{eq:theta_2_code} whose repeated
application to a particular super-star ``1'' is equivalent to a Monte
Carlo integration of the DCs\footnote{As the super-star is moved around
  on its orbit between two numerical encounters, the procedure is best
  described as an implicit evaluation of the \emph{orbit-averaged}
  DCs.}, provided the properties of field particles ``2'' are
correctly sampled. Under the usual assumption that encounters are
local, this latter constraint is obeyed if we take these properties to
be those of the closest neighboring super-star.  Furthermore, this
allows us to actually modify the velocities of both super-stars at a
time, each one acting as a representative from the ``field'' for the
other. Hence, in the H\'enon code (as well as in ours), super-stars are
evolved in symmetrical pairs. This does not only speed up the
simulations by a factor $\simeq 2$, but also ensures proper local
conservation of energy, a feature which turned out to be a 
prerequisite for correct cluster evolution. Unfortunately this
pairwise approach also impose heavy constraints on the code's
structure and (perhaps) abilities as we shall show later on.

So the elementary ingredients in the heart of H\'enon's scheme are
simulated 2-body gravitational deflections between neighboring
super-stars. However, instead of being direct one-to-one counterparts
to real individual encounters -- which would lead to much too slow a
code with a (huge) number of computational steps scaling as
$N_\mathrm{part}^2$ -- these are actually ``super-encounters'',
devised to statistically reproduce the cumulative effects of the
numerous physical deflections taking place in the real system over a
time span $\delta t$. Thus, such a numerical encounter has a double
nature. First, it is computed as a (virtual) 2-body gravitational
interaction with deflection angle $\theta_\mathrm{SE}$ in the pair CM
frame. But being in charge of representing all the (small-angle)
deflections that test-star ``1'' experiences during $\delta t$ when
meeting field-stars ``2'', it also has to obey
Eqs.~\ref{eq:theta_2_code2},~\ref{eq:trel_encounter}. Consequently,
$\theta_\mathrm{SE}$ has to equate the root mean squared cumulative
deflection
\begin{equation}
  \theta_\mathrm{SE} = \frac{\pi}{2}\sqrt{\frac{\delta t} 
    {\widehat{T}_\mathrm{rel}^{(1,2)}}}.
  \label{eq:theta_SE}
\end{equation}

This double nature of the encounter is reflected in the whole MC
scheme that can be regarded either, quite abstractly, as a stochastic
algorithm to solve the Fokker-Planck equation or, more simply, as some
kind of randomized N-body scheme. The second viewpoint, though it
might be misleading on certain occasions, is the one we usually adopt
as it allows more intuitive reasoning.

We now describe the computation of a particular numerical encounter.
It decomposes into the following steps:
\begin{enumerate}
  \setcounter{enumi}{-1}
\item A pair of adjacent super-stars and a time step $\delta t$ are
  chosen by a procedure to be explained in
  Sect.~\ref{sec:Rel_Pair_Choice}.
\item The local density $n$ entering the determination of
  $\widehat{T}_\mathrm{rel}^{(1,2)}$ in Eq.~\ref{eq:theta_SE} is
  estimated.
\item The super-stars' velocities, $\vec{v}_1$ and $\vec{v}_2$ are
  randomly oriented while respecting the angular momenta
  $J_i=\|\vec{J_i}\|$ and specific kinetic energy
  $T_i=\frac{1}{2}\vec{v_i}^2$ of both super-stars. This sets the CM-
  and relative velocities $\vec{v}_\mathrm{CM}$ and
  $\vec{v}_\mathrm{rel}$. The former defines the encounter CM frame
  while the latter allows $\theta_\mathrm{SE}$ to be determined through
  Eqs.~\ref{eq:trel_encounter} and \ref{eq:theta_SE}.
\item In the CM frame, the orientation of the orbital plane is
  randomly chosen around the direction of $\vec{v}_\mathrm{rel}$.
  $\theta_\mathrm{SE}$ being known, computing the post-encounter
  velocities in the CM frame is trivial.
\item These velocities are transformed back to the cluster frame where 
  they define new $J_i$s and $T_i$s for both super-stars.
\end{enumerate}

To compute the local density of star $n(R)$ required in step 1, we
build and maintain a radial ``Lagrangian'' mesh each of whose cells
typically contains a few tens of super-stars.\footnote{The mesh
  is entirely rebuilt when the number of super-stars in any of its
  cells deviates from some prescribed interval, typically $[12, 36]$.
  In that way, the mesh structure always matches quite closely the
  super-star distribution.} The cells' radial limits are known, as
well as the number of super-star they contain. Hence, an estimate of
the local number density is easily computed by dividing the total
number of stars in the cell where the encounter takes place by its
volume.  Frequent updating (after each super-star orbital movement)
and occasional rebuilding of the mesh introduces only a very slight
computational overhead, most CPU time being spent in binary tree
traversals during potential and rank computations (see
Sect.~\ref{sec:pot_rep}).

\modif[10]{The computations in steps 2-4 are described in
appendix~\ref{app:bin_tree}.} The only physical content of all this
procedure resides in the determination of $\theta_\mathrm{SE}$.
Everything else is a matter of elementary frame transformations and
correct random sampling of free parameters in the Monte Carlo spirit.

\subsubsection{Choice of the relaxation pair. Determination of the
  time step.}
\label{sec:Rel_Pair_Choice}

With no other physical process than relaxation included, each
individual step in our algorithm comprises three operations:
\begin{enumerate}
  \item Selection of a pair of neighboring super-stars to be evolved.
  \item Modification of their orbital properties ($E_i$ and $J_i$) through
    a super-encounter, as explained above.
  \item For each super-star, selection of a new position on the
    ($E_i^\prime$,$J_i^\prime$)-orbit, i.e. determination of a new
    radius $R_i$ for the spherical shell. This comprises an adequate
    update of the cluster's potential.
\end{enumerate}
At that point, the code cycles back to 1, i.e., another pair is chosen
and another step begins. This crucial selection process is presented
in this section.

The choice procedure is mainly constrained by the necessity of
allowing super-stars to have individual time-steps $\delta t_i$ that
reflect the enormous variations of the relaxation time between the
central and outer parts of the stellar cluster. When collisions are
included, the dynamical range of evolution times can be even
larger. Unless this specification is met, the code's efficiency would
be very low as the overall $\delta t$ would have to reflect the very
short central evolution time. One could also be concerned by the
orbital time exceeding $\delta t$ for a large fraction of super-stars,
a situation inconsistent with the ``orbital average'' approach
implicitly assumed in the Monte Carlo scheme. However, this problem is
actually nonexistent in a purely relaxational system whose evolution
-- under the assumptions made in the standard relaxation theory -- is
independent of $T_\mathrm{relax}/T_\mathrm{dyn}$, provided $N_{\ast}$
is large enough ($\gg 100$).

The other important constraint is the need to evolve both super-stars
in an interacting pair. If the same time step is not used for both
super-stars, energy is not conserved and a very poor cluster simulation
ensues.
But adjacent super-stars only form a pair during a unique interaction
and then break apart as each one is attributed a new radius. So,
momentary neighboring super-stars have to be given similar $\delta
t_i$. This strongly suggests use of \emph{local} time steps, i.e.
$\delta t$ should be a function of $R$ alone. 

Naturally, the time steps have to be sufficiently smaller than the
time scale for the physical processes driving the cluster evolution,
namely the relaxation in the present case. Hence, we impose:
\begin{equation}
  \delta t(R) \leq \delta t_\mathrm{opt}(R) = f_{\delta t} \tilde{T}_\mathrm{rel}(R)
  \label{eq:ineq_dt_rel}
\end{equation}
where $\tilde{T}_\mathrm{rel}$ is some kind of locally averaged
relaxation time defined as (see Eq.~\ref{eq:trel_encounter}):
\begin{equation}  
  \tilde{T}_\mathrm{rel} \propto
  \frac{\langle v^2 \rangle^\frac{3}{2}}
  {\ln\left(\gamma N_\ast\right) G^2 n \langle M_\ast \rangle^2}
  \label{eq:Trel_loc}
\end{equation}
and $f_{\delta t}=0.005-0.05$ typically.

In Eq.~\ref{eq:Trel_loc}, $n$ (star number density), $\langle v^2
\rangle$ (average squared velocity) and $\langle M_\ast \rangle$
(average stellar mass) are $R$-dependent properties of the cluster. As
the only role of $\tilde{T}_\mathrm{rel}$ is to provide short enough
$\delta t_i$s, an approximate evaluation of these quantities (using a
coarse mesh or a sliding average) is sufficient. On the other hand,
too short $\delta t_i$s would fruitlessly slow down the code and
should be avoided by considering $\delta t_\mathrm{opt}$ in
Eq.~\ref{eq:ineq_dt_rel} not only as an upper limit for the time step
but also as an optimal value to be approached as closely as possible.

As the members of a pair arrived at their present position at
different times but have to leave it at the same time, once the
super-encounter is performed, imposing the same $\delta t$ to both
super-stars is impossible. So, building on the statistical nature of the
scheme, instead of trying to maintain a super-star at radius $R$
during exactly $\delta t(R)$, we only require the \emph{mean} waiting
time for super-stars at $R$ to be $\delta t(R)$. As explained by
\citet{Henon73}, this constraint is fulfilled if the probability for a
pair lying at $R$ to be selected and evolved (and thus, taken away
from $R$) during a time span $dt$ is $P_\mathrm{selec}(R) = dt/\delta
t(R)$. This is realised in the following way:
\begin{itemize}
\item As it would be difficult to define and use a selection
  probability $P_\mathrm{selec}$ which is a function of the continuous
  variable $R$, we define it to depend on the rank $i$
  of the pair (rank 1 designates the two super-stars that are closest to
  the centre, rank 2 the second and third super-stars, in ascending
  order of $R$ and so on). For a given
  cluster's state, local relaxation times $\tilde{T}_\mathrm{rel}$ are
  computed at the radial position of every pair. Rank-depending time
  steps are defined that obey inequality~\ref{eq:ineq_dt_rel},
  \begin{equation}
    \delta t(i) \le f_{\delta t} 
    \tilde{T}_\mathrm{rel}(R(i)).
  \end{equation}
\item Normalized selection probabilities are computed,  
  \begin{equation}
    P_\mathrm{selec}(i) = \frac{\overline{\delta t}}{\delta
      t(i)}
    \mbox{\ \ with\ \ } \overline{\delta t} = \left(
    \sum_{j=1}^{N_\mathrm{SS}-1} \frac{1}{\delta t(j)} \right)^{-1} 
  \end{equation}
  from which we derive a cumulative probability,
  \begin{equation}
    Q_\mathrm{selec}(i) =
    \sum_{j=1}^{i} P_\mathrm{selec}(j).
  \end{equation}
\item At each evolution step another super-star pair is randomly chosen
  according to $P_\mathrm{selec}$. To do this, a random number
  $X_{\mathrm{rand}}$ is
  first generated with uniform probability between 0 and 1. The pair
  rank is then determined by inversion of
  $Q_\mathrm{selec}$:
  \begin{equation}
    i = Q_\mathrm{selec}^{-1}(X_{\mathrm{rand}}).
  \end{equation}
  The binary tree (See Sect.~\ref{sec:pot_rep}) is traversed twice to
  find the id-numbers of the member super-stars whose (momentary)
  ranks are $i$ and $i+1$.
\item The pair is evolved through a super-encounter, as explained in
  Sect.~\ref{sec:super_encounter}, for a time step $\delta
  t(i)$.
\item After a large number of elementary steps, typically
  $0.5\,N_{\mathrm{SS}}$, the $\delta t(i)$ and
  $P_\mathrm{selec}(i)$ are re-computed from scratch to
  reflect the slight modification of the overall cluster structure.
\end{itemize}

\begin{figure*}
  \resizebox{12cm}{!}{
    \includegraphics{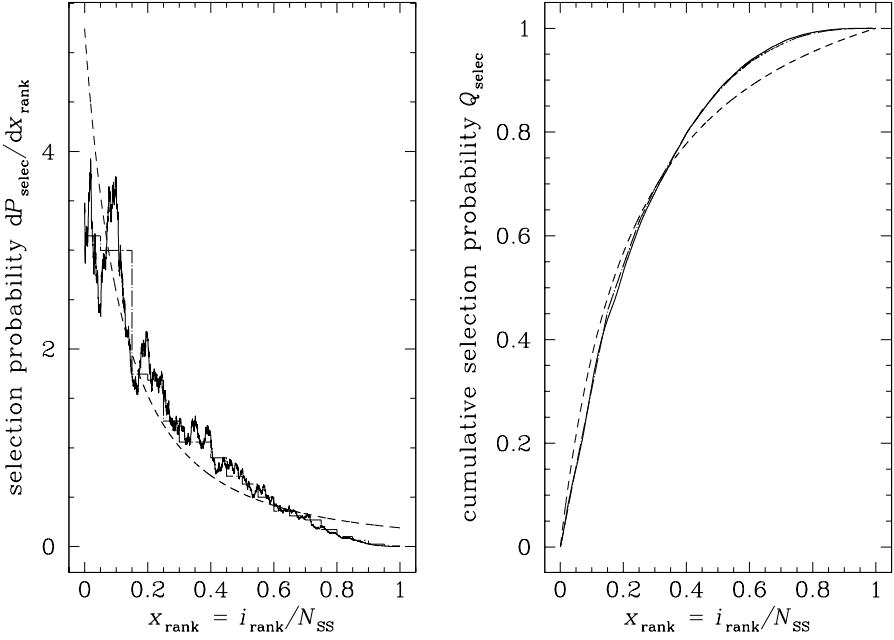}% Pselec_Pselec_cum_plummer.eps --> MS1125f1.eps
    }
  \hfill
  \parbox[b]{55mm}{
    \caption{Selection probability in a Plummer cluster consisting of
      4000 super-stars. Solid curves correspond to $P_\mathrm{selec}
      \propto 1/T_\mathrm{rel}$, dashed ones to H\'enon's recipe and
      dot-dashed ones to our piecewise approximation.}
    \label{fig:pselec}
    }
\end{figure*}

As evaporation, collisions and tidal disruptions remove stars from
the cluster, the number of super-stars $N_\mathrm{SS}$ generally
decreases between two successive computations of the selection
probabilities. To avoid this problem, $\delta t$, $P_\mathrm{selec}$ and
$Q_\mathrm{selec}$ are actually defined as functions of
$x_\mathrm{rank}=i/N_\mathrm{SS} \in ]0,1]$.

For the sake of efficiency, we should manage to get for 
$Q_\mathrm{selec}^{-1}$ a function that is quickly evaluated. We
explored two solutions. We first mimicked H\'enon's recipe, using the
functional forms:

\begin{eqnarray}
  P_\mathrm{selec}(i) &=& \frac{ (1+C)CN_\mathrm{SS} }{
    (i+CN_\mathrm{SS}) (i+CN_\mathrm{SS}-1) } 
  \label{eq:Pselec_Henon} \\
  Q_\mathrm{selec}(i) &=& \frac{ (1+C)i }{ i+CN_\mathrm{SS} } 
  \label{eq:Qselec_Henon} \\
  Q_\mathrm{selec}^{-1}(x) &=& \left\lfloor 1 + \frac{ CN_\mathrm{SS}
      }{ 1+C-x } \right\rfloor  
  \label{eq:Qinv_selec_Henon} \\
  \delta t(i) &=& \frac{\overline{\delta t}}{P_\mathrm{selec}(i)}.  
  \label{eq:dt_from_Pselec} 
\end{eqnarray}
The parameter $C > 0$ controls the probability contrast between the
centre (short $T_\mathrm{rel}$ and $\delta t$) the outer regions (long
$T_\mathrm{rel}$ and $\delta t$). The lower the value of $C$, the more
centrally peaked is $P_\mathrm{selec}$. We determine $C$ by
least-square fitting $Q_\mathrm{selec}(i)$ to $Q_\mathrm{rel}(i)
\propto \sum_{j=1}^i T_\mathrm{rel}^{-1}(j)$\footnote{In the least
  square procedure, points are weighted by $T_\mathrm{rel}^{-1}(j)$ in
  order to get a better agreement for the most frequently selected
  ranks}.

However, being rather arbitrary, H\'enon's parameterization could lead
to values of $\delta t$ that poorly match the shape of
$T_\mathrm{rel}(R)$ with the effect of forcing low $\overline{\delta
  t}$ and hence slowing down the simulation. This concern motivated
another approach. The full rank range is sliced in a few (typically
20) bins, in each of which the selection probability is set to a
constant (either proportional to the maximum of $T_\mathrm{rel}^{-1}$
in the bin or to its mean value). Such a piecewise approximation is
naturally expected to adjust better to any cluster structure, as shown
in Fig.~\ref{fig:pselec}.

An additional constraint to be mentioned is the need to restrict the
ratio of the longest time step to the shortest one, $\delta
t_\mathrm{max} / \delta t_\mathrm{min}$, to ensure that the outermost
shells (the cluster's halo) evolves correctly. Otherwise these
super-stars, most probably placed near their apo-centre positions,
where relaxation (and, hence selection probability) is vanishing, would
never be given any opportunity to visit more central regions where
they can react to the adiabatic relaxation-induced modification of the
central potential and experience 2-body encounters\footnote{If
  super-stars could be attributed individual time steps, setting
  $\delta t^{-1}$ to an orbital average of $T_\mathrm{rel}^{-1}$
  would naturally prevent super-stars with small enough
  $R_\mathrm{p}$ from ``freezing'' in the cluster's outskirts. Our
  procedure amounts to such an orbital averaging, in a Monte Carlo
  fashion. It will fail if the time lag between two successive
  selections of a given super-star is not small enough compared with
  the time over which substantial alterations of the cluster's
  structure occur.}. Finally, for reasons to be presented in
appendix~\ref{app:placement}, it is necessary that the selection
probability is a decreasing function of $R$ (i.e. of the rank). In
practice, these added constraints are applied to modify unnormalised
selection probabilities which are then rescaled to 1. Once the
probabilities have been worked out, we ensure
inequality~\ref{eq:ineq_dt_rel} is satisfied everywhere by setting
\begin{equation}
  \overline{\delta t} = f_{\delta t} \,\max(T_\mathrm{rel}(i)\cdot 
  P_\mathrm{selec}(i)),
\label{eq:delta_t_bar}
\end{equation}
where the maximum is taken over all super-stars.

\begin{figure}
  \centering
  \resizebox{0.95\hsize}{!}{
    \includegraphics{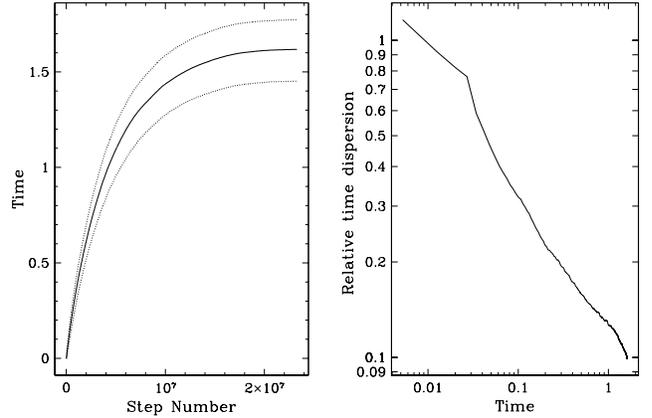}}% DeltaT_code_MC.eps --> MS1125f2.eps
  \caption{Evolution of the time dispersion of super-stars during the
    simulation of a core-collapsing cluster. Left panel: Time
    versus step number. The median time $t_\mathrm{med}$ is shown as a
    solid line. Dotted lines depict the interval
    $[t_\mathrm{med}-\Delta t_{-};t_\mathrm{med}+\Delta t_{+}]$
    comprising $2/3$ of super-stars. Right Panel: time evolution of
    the relative time ``dispersion'' $\Delta t/t_\mathrm{med}$ with
    $\Delta t = 0.5(\Delta t_{-}+\Delta t_{+})$.}
  \label{fig:EvolT_DeltaT}
\end{figure}

As super-stars are randomly chosen to be evolved and advanced in time,
strict synchronization is never realized (except at $t=0$). Each
super-star $k$ has its own individual time $t^{(k)}$ and
synchronization is only achieved \emph{statistically} by requiring
that, at every stage in the cluster's evolution, the expectation
values $E(t^{(k)})$ of all $t^{(k)}$s are the same. An equivalent
statement is to impose an equal expectation value
$E_\mathrm{step}\left(\delta t^{(k)}\right)$ for the individual time
increase of any super-star during any evolution step. At the beginning
of a given step, the super-stars are ranked according to their distance
to the centre. Selection probability and time step depend only on the
rank number $i(k)$ so that
\begin{equation}
  E_\mathrm{step}\left(\delta t^{(k)}\right) = 
  \underbrace{ P_\mathrm{selec}(i) }_{ 
    \parbox{2cm}{\center\scriptsize 
      \vspace{-0.3cm}
      Probability for selecting the super-star with rank \(i\).
      }
    }
  \times  
  \underbrace{\delta t(i) }_{ 
    \parbox{2cm}{\center\scriptsize
      \vspace{-0.3cm}
      Time step for this rank.
      }
    }
\end{equation}
and Eq.~\ref{eq:dt_from_Pselec} yields:
\begin{equation}
  E_\mathrm{step}\left(\delta t^{(k)}\right) = P_\mathrm{selec}(i)  
  \frac{\overline{\delta
      t}}{P_\mathrm{selec}(i)} = \overline{\delta t}~~\forall k
\end{equation}
as desired. Figure~\ref{fig:EvolT_DeltaT} illustrates how the dispersion
of super-stars' times evolves during a typical simulation. \modif[10]{We define
the global, ``cluster'' time to be the median value of the
super-stars' times.}

\section{Orbital displacements and potential updating.}
\label{sec:orbit}

\subsection{Potential representation}
\label{sec:pot_rep}

In Sect.~\ref{sec:std_relax_theory}, we explain how relaxation theory,
as adopted in this work, relies on the assumption that the cluster's
gravitational potential $\Phi$ can be described as a dominating smooth
contribution $\Phi_\mathrm{s}$ whose evolution time scale is much
longer than the typical orbital time plus a relatively small
fluctuating $\delta\Phi$. This latter contribution being further
reduced to the sum of numerous 2-body encounters, we are left with the
numerical representation of $\Phi_\mathrm{s}$.

As spherical symmetry introduces many simplifications, going beyond
this central approximation deeply built into H\'enon's scheme, seems
nearly impossible. Its most prominent merit is to ensure that stellar
orbits, when considered on time scales much shorter that
$T_\mathrm{relax}$, are easily dealt-with planar \emph{rosettes}.
Therefore, angular $\Phi$-fluctuations are removed by construction and
we represent $\Phi_\mathrm{s}$ as the sum of the contributions of the
super-stars, i.e., spherical infinitely thin shells of stars. As a
consequence, radial graininess is still present but its effect turns
out to be insignificant compared to ``genuine''
relaxation\footnote{Moreover, unlike physical relaxation which only
  depends on the simulated number of stars $N_\ast$, radial
  ``numerical'' relaxation vanishes as we increase resolution (i.e.
  the number of super-stars $N_\mathrm{SS}$), see
  appendix~\ref{app:spur_rel}.}. To support this claim, we switched
off simulated physical relaxation and got a cluster showing no sign of
evolution (apart from Monte Carlo noise) for a number of steps at
least three times larger than needed to accomplish deep core collapse
when relaxation is included.

Between two successive super-stars of rank $i$ and $i+1$, the smooth
potential felt by a thin shell of mass $M$ with radius $R 
\in [R_i,R_{i+1}]$ is then simply 
\begin{eqnarray}
  \lefteqn{ \Phi_\mathrm{s}(R) = -\frac{A_i-0.5M}{R} - B_i }&&\\
  & & \mbox{with}~ 
  A_i=M_\mathrm{BH}+\sum_{j=1}^{i-1} M_j ~\mbox{and}~ 
  B_i=\sum_{j=i}^{N_\mathrm{SS}} \frac{M_j}{R_j} \nonumber
\end{eqnarray}
where $M_j$ and $R_j$ are the mass and radius of the super-star of
rank $j$ and $M_\mathrm{BH}$ is the mass of the central BH. The term
$0.5M/R$ is due to shell self-gravitation. To lighten notations,
from this point on, $\Phi_\mathrm{s}$ will simply be referred to as the
``potential'' and the symbol $\Phi$ is re-attributed to it.

At each step in the numerical simulation, two super-stars are evolved
which are given new radii and masses (if collisions or stellar
evolution is simulated). An easy way of obtaining exact overall energy
conservation and proper account of the adiabatic energy drift of
super-stars is to update the $A_i$ and $B_i$ coefficients after every
such orbital displacement. Doing so also ensures that we never put a
super-star at a radius which turns out to be forbidden (either lower
than peri-centre or larger than apo-centre) in the updated potential.
To sum it up, this choice spares us much trouble connected with a
potential that lags behind the super-star distribution.
\citet{Stodol82} and \citet{Giersz98} describe these difficulties, as
well as recipes to overcome them in their programs. Similar problems
are certainly present in the code of \citet{JRPZ00} as they recompute
the potential only after all the super-stars have been assigned new
radii. However, performing potential updates only after a large number
of super-star moves has advantages of its own. In particular, in such
a code, the computing time should scale linearly with the number of
super-stars (for a complete cluster evolution). This also allows them
to develop parallelized versions of the Monte Carlo scheme
\citep{JRPZ00}.

If we implement the $A_i$s and $B_i$s as linear arrays, however, a
large fraction of their $N_\mathrm{SS}$ elements would have to be
modified after \emph{each} step, so the number of numerical operations
required to evolve the system to a given physical time would scale
like $N_\mathrm{SS}^2$. This steep dependency could be avoided by
using a {\bf binary tree} data structure to store the potential (and
ranking) information \citep{Sedgewick88}. This essential adaptation
was alluded to by H\'enon himself \citep{Henon73} who never published
it though.

At any given time, each super-star is represented by a node in the
tree.  Each node is connected to (at most) two other nodes that we
shall call his left and right children, which are themselves, when
present, the ``roots'' of their father's left and right ``sub-trees''.
The rules underlying the tree structure are that {\bf all the nodes in
  the left ``child-tree'' of a given node correspond to super-stars
  with lower radii} and {\bf all the nodes in its right ``child-tree''
  to super-stars with higher radii}. If we define $\mathcal{LT}_k$ and
$\mathcal{RT}_k$ to be the sets of nodes in the left and right
child-trees of node $k$, then
\begin{equation}
  \label{eq:btree_R_rule}
  \begin{minipage}{6.8cm}
    \vskip -5mm
    \begin{eqnarray}
      R_k &\ge& R_m\;\; \forall m \in \mathcal{LT}_k  \nonumber \\
      R_k &<  & R_m\;\; \forall m \in \mathcal{RT}_k. \nonumber 
    \end{eqnarray}
  \end{minipage}
\end{equation}

\begin{figure}
  \resizebox{\hsize}{!}{
    \includegraphics{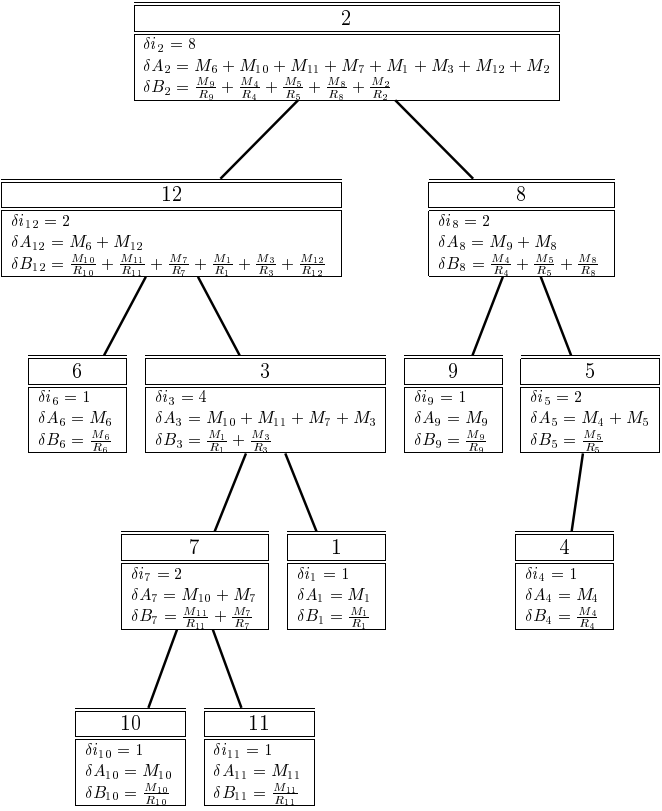}% bin_tree2.eps --> MS1125f3.eps
    }
  \caption{Diagram of a binary tree with 12 super-stars showing node
    properties that encode potential ($\delta A_k$, $\delta B_k$)
    and rank ($\delta i_k$) information.}
  \label{fig:bin_tree2}
\end{figure}

The spherical potential is represented by $\delta A_k$ and $\delta
B_k$ coefficients attached to nodes. A third value, $\delta i_k$,
allows the determination of the radial rank of any super-star. These
properties, illustrated in Fig.~\ref{fig:bin_tree2}, are defined by
\begin{equation}
  \label{eq:btree_dAdB_def}
  \begin{minipage}{6.8cm}
    \vskip -5mm
    \begin{eqnarray}
      \delta i_k &=& 1 + 
      \mbox{number of nodes in~} \mathcal{LT}_k,    \nonumber \\
      \delta A_k &=& M_k + 
      \sum_{m \in \mathcal{LT}_k} M_m \mbox{~~and}  \nonumber \\
      \delta B_k &=& \frac{M_k}{R_k} + 
      \sum_{m \in \mathcal{RT}_k} \frac{M_m}{R_m}. \nonumber 
    \end{eqnarray}
  \end{minipage}
\end{equation}
\modif[11]{Then, the super-star with rank $i$ can be found and its
  potential coefficients $A_i$ and $B_i$ can be computed by traversing
  the tree from root to the node associated with this particle. During
  the same traversal, the $\delta A_k$s, $\delta B_k$s and $\delta
  i_k$s of the visited nodes may be modified in order to account for
  the removal of this super-star. After the super-star has been given
  a new radius (see Sect.~\ref{sec:selecRorb}), it is re-introduced
  into the tree through another traversal. Hence, the potential and
  rank coefficients reflect always exactly the positions (and masses)
  of all the super-stars.  The potential at any radius, as well as the
  peri- and apo-centre distances of a given super-star, can be computed
  by specific tree traversal routines.  The number of operations
  involved in any of these tree traversals is proportional to the
  number of levels i.e., if the tree is kept reasonably well balanced,
  $\mathcal{O}(\log_2(N_\mathrm{SS}))$. From time to time, (typically
  after every super-star has been evolved 2 times on average), the
  binary tree is rebuilt from scratch, in order to keep it well
  balanced and to remove empty nodes.  More details about the
  implementation of this binary tree can be found in
  appendix~\ref{app:bin_tree}.}

\subsection{Selection of a new orbital position}
\label{sec:selecRorb}

Let's consider a star with known energy $E$ and angular momentum $J$
orbiting in a spherically symmetric potential $\Phi(R)$. Its distance
to the centre $R$ oscillates between $R_\mathrm{p}$ (peri-centre
radius) and $R_\mathrm{a}$ (apo-centre radius) which are the two solutions of
the equation
\begin{equation}
  \label{eq:vrad2eq0}
  v_\mathrm{rad}^2 = 2E-2\Phi(R)-\frac{J^2}{R^2} = 0.
\end{equation}
During a complete orbit\footnote{As orbits are generally not closed
  curves but rosettes, a ``complete orbit'' is defined as the segment
  of trajectory between successive passages at the peri-centre.} the
time spent in radius interval $[R,R+\mathrm{d}R]$ is
$\mathrm{d}t \propto v_\mathrm{rad}^{-1}(R) \mathrm{d}R$ so that the
probability density for finding the star at $R$ at any random time (or
at a given time but without any knowledge about the orbital phase) is
\begin{equation}
  \label{eq:dPorbdR}
  \frac{\mathrm{d}P_\mathrm{orb}}{\mathrm{d}R} = \frac{2}{P_\mathrm{orb}} 
  \frac{1}{v_\mathrm{rad}(R)}
\end{equation}
where $P_\mathrm{orb} = 2 \int_{R_\mathrm{p}}^{R_\mathrm{a}}
\mathrm{d}R\, v_\mathrm{rad}^{-1}$ is the orbital period.

Our Monte Carlo scheme avoids explicit computation of the orbital
motion. % Necessiterait une explication theorique mieux fondee
It instead achieves correct statistical sampling of the orbit of any
given super-star by ensuring that the expectation value for the
fraction of time spent at $R$ complies with Eq.~\ref{eq:dPorbdR}. Let
the sought-for probability of placing the super-star at $R \in
[R_\mathrm{p},R_\mathrm{a}]$ be $f_\mathrm{plac}(R) =
\mathrm{d}P_\mathrm{plac} / \mathrm{d}R$. According to
Eq.~\ref{eq:dt_from_Pselec}, if the super-star is placed at $R$, it will
stay there for an average time $\overline{\delta
  t}/P_\mathrm{selec}(R)$. Then, combining both relations, the average
ratio of times spent at two different radii $R_1,R_2$ on the orbit is
\begin{eqnarray}
  \left\langle \frac{t_\mathrm{stay}(R_1)}{t_\mathrm{stay}(R_2)}
  \right\rangle & = &
  \frac{ 
    f_\mathrm{plac}(R_1)P_\mathrm{selec}(R_2) }{ 
    f_\mathrm{plac}(R_2)P_\mathrm{selec}(R_1) 
  } \\ 
& = & 
  \frac{ 
    v_\mathrm{rad}(R_2) }{ 
    v_\mathrm{rad}(R_1) 
  } \mbox{~~ as required by Eq.~\ref{eq:dPorbdR}.}
\end{eqnarray}
As a result, this imposes
\begin{equation}
  \label{eq:dPplacdR}
  f_\mathrm{plac}(R) \propto \frac{ P_\mathrm{selec}(R) }{
    v_\mathrm{rad}(R) }.
\end{equation}
%$

\modif[11]{In appendix~\ref{app:placement}, we explain how we
  implement such a probability function in an efficient way.}

\section{Other ingredients}
\label{sec:other}

\subsection{Evaporation and tidal truncation}

Despite its long history, the theoretical understanding of the
processes leading stars to escape a stellar cluster is still not
complete (\citealt{BT87} Sect.~8.4.1; \citealt{MH97} Sect.~7.3;
\citealt{Heggie00}). Even without considering interaction with
binaries, the global picture seems a bit confusing.  Nevertheless,
\modif[12]{for an isolated cluster}, the basic mechanism is obvious to
grasp, at a ``microscopic'' description level: a star can escape after
it has experienced a 2-body encounter which resulted in an energy gain
large enough to unbound it, i.e., \modif[12]{to get to positive
  energy}. Much of the confusion about the prediction of overall
escape rates amounts to figuring out whether rare large angle
scatterings that are neglected by the standard relaxation theory could
dominate this rate. Indeed it can be argued that, in an isolated
cluster, stars that are only slightly bound and could be kicked away
by weak scatterings populate orbits with huge periods and spend most
time near the apo-centre where encounters are vanishingly rare
\citep{Henon60}.  According to that picture, the escape rate could not
be predicted by the ``standard'' relaxation theory, because individual
``not-so-small'' angle scatterings would dominate it.

If this is true, as the MC treatment of 2-body encounters relies on
the assumption of small relative changes in orbital parameters, the
method cannot be expected to give reliable results for the escape rate
from an isolated cluster \citep{Henon71b}.  Some numerical solution to
that problem was introduced by \citet{Giersz98}.  \modif[12]{However,
  when the cluster's initial conditions are set to represent a
  galactic nucleus, the fraction of stars that evaporate during
  $10^{10}$~years is very small, so that a precise account of this
  phenomenon is not really required. It would anyway not make much
  sense to devise a complicated treatment of evaporation from the
  nucleus while we neglect the inverse process, i.e. the capture of
  stars from the galactic bulge. Our procedure is simply to remove any
  star whose energy is positive after a relaxation/collision process.
  As can be seen in Fig.~\ref{fig:plum_comp_giersz}, this simple
  prescription leads to an amount of evaporation in good agreement
  with the result of \citet{Giersz98}.}

However, for the sake of comparison with globular cluster
simulations, we also introduced the effects of an external (galactic)
tidal field.  Due to the sphericity constraint, the three dimensional
nature of such a perturbation cannot be respected. We resort to the
usual radial truncation approach and consider that a super-star with
apo-centre radius larger than the so-called tidal radius
$R_\mathrm{tid}$ immediately leaves the cluster\footnote{Actually, the
  star would still wander through the cluster for a period $\sim
  P_\mathrm{orb}$ which could be an appreciable fraction of the
  cluster evolution time scale for low $N_\ast$. Neglecting that fact
  could lead to a strong disagreement between $N$-body and
  Fokker-Planck based models \citep{TPZ98}. In fact, as the star has
  to find the ``exit door'' near the Lagrange points before it can
  effectively escape, it may stay in the cluster for many dynamical
  times even though its energy is well above the escape energy.
  Globular clusters may thus contain a large amount of ``potential
  escapers'' \citep{FH99,Baumgardt00}.}.  The value of
$R_\mathrm{tid}$ is about the size of the cluster's Roche lobe,
$R_\mathrm{tid} = c R_\mathrm{gal}
(M_\mathrm{cl}/M_\mathrm{gal})^{1/3}$ with $R_\mathrm{gal}$ being the
distance to the centre of the parent galaxy whose mass is
$M_\mathrm{gal}$ and $c$ is of order unity. \modif[13]{This criterion
  is clearly a quite unrealistic simplification but we do not question
  it in our work as it is used only for comparison purposes.  As the
  transition from an apo-centre radius slightly below $R_\mathrm{tid}$
  to a value slightly larger does not imply large changes in the shape
  of the orbit, the escape rate in this model is certainly dominated
  by small angle, diffusion-like relaxation and must be correctly
  captured by our MC approach.}

\subsection{Neglect of binary processes}
\label{sec:binaries}

The formation, evolution and dynamical role of binaries in star
clusters are complex and fascinating subjects.  An impressive number
of works have been aimed at the study of binaries in globular clusters
\citep[see, for instance][ for a review]{Hut92}.  On the other hand,
not much has been done concerning galactic nuclei
\citep[see][]{Gerhard94}.

From a dynamical point of view, only \emph{hard} binaries, i.e. star
couples whose orbital velocity $v_\mathrm{orb}$ is larger than the
velocity dispersion $\sigma_v$ in the cluster, have to be considered.
In dense systems, they act as a heat source by giving up orbital
energy and contract (thus hardening further) during interactions with
other stars. Of course, the fraction of primordial binaries to be
labeled as ``hard'' is \emph{a priori} much higher in globular
clusters ($\sigma_v$ of order $10\,\mathrm{km\,s}^{-1}$) than
in galactic nuclei ($\sigma_v \ge 100\,\mathrm{km\,s}^{-1}$).
Binaries can also be formed dynamically, either by tidal energy
dissipation during a close 2-body encounter (``tidal binaries'') or as
the result of the gravitational interaction between 3 stars (``3-body
binaries''). The cross section for forming a tidal binary strongly
decreases with relative velocity (at infinity) $v_\mathrm{rel}$
\citep{KL99}, so that, in galactic nuclei, such processes imply
hydrodynamic contact interactions that are likely to result in mergers
\citep{LN88,BH92,LRS93}. Thus these events are implicitly treated in
our code as a subset of all the collisions \citep{FB01b}.  An
interesting counter-example to which these arguments do not apply is
the nucleus of the nearby spiral galaxy M33 whose central velocity
dispersion is as low as $\sim 20\,\mathrm{km\,s}^{-1}$
\citep{Lauer98} so that tidal binaries should have formed at an
appreciable rate \citep{HHK91}.

The formation rate of 3-body binaries in galactic nuclei is also
strongly quenched as compared to globular clusters. Indeed, for a
self-gravitating system, the total number of binaries formed through
this mechanism per relaxation time is only of the order of
$N_\mathrm{3bb} \sim 0.1/(\ln\Lambda\,N_\ast)$ \citep{BT87,GH93} and
can be completely neglected unless evolution, through mass-segregation
and core collapse, leads to the formation of a dense auto-gravitating
core containing only a few tens of stars \citep{Lee95}.  Finally,
another, somewhat exotic, possibility is the formation of hard
binaries by radiative energy losses of gravitational waves during
close fly-bys between two compact stars \citep[ for
instance]{LeeMH93}. Note that, if present, hard binaries would not
only have a dynamical role but may also destroy giant stars by
colliding with them \citep{DBBS98}.
  
For these reasons, we feel justified not to embark on the considerable
burden that incorporation of binary processes in a Monte Carlo scheme
would necessitate. However, this has been achieved with a high level
of realism by \citet{Stodol86}, Giersz \citeyearpar{Giersz98,Giersz00b}.
Such a detailed approach is required to obtain reliable rates for
binary processes of interest, like super-giant destruction by
encounters with binaries \citep{DBBS98}, but, if needed, the basic
dynamical effect of binaries as a heat source could be accounted for
much more easily using the same recipes that proved suitable in direct
Fokker-Planck methods \citep[ for instance]{LFR91}.

It should be noted that, even in the absence of any explicit
simulation of binary heating, core collapse is anyway halted and
reversed in most Monte Carlo simulations of globular cluster
evolution! This is due to an effect already described by
\citet{Henon75} and \cite{Stodol82}: a stiff micro-core consisting of
one or a few ($\le 5$) super-stars becomes self-gravitating and
misleadingly mimics a small set of hard binaries by contracting and
giving up energy to other super-stars. Due to the self regulating nature
of cluster re-expansion \citep{Goodman89}, this leads to a
post-collapse evolution of the overall cluster structure that is
extremely similar to what binaries produce.  Unfortunately this does
not hold true for the very core whose evolution (for instance, whether
it experiences gravothermal oscillations or not, see
\citealt{Heggie94}) depends on the nature of the heat source.
 
\section{Code testing}
\label{sec:tests}

In this section, we briefly describe the results of a series of test
simulations we conducted in order to check the various aspects of 
our code and the results it produces.

\subsection{Dynamical equilibrium \& spurious relaxation}

The most basic test to be passed is to make sure that when relaxation
and other physical processes are turned off, no evolution occurs in a
cluster model whose initial conditions obey dynamical equilibrium and
radial stability.  \modif[15]{Beyond stability by itself, the main concern is
about {\em spurious relaxation} introduced by the discrete
representation of the cluster by a set of super-stars. In other words,
the supposedly ``smooth'' potential $\Phi$ still presents radial
graininess that could induce some kind of unwanted relaxation
\citep{Henon71b}. It can easily be shown that the time scale over
which the effects of this spurious relaxation may become of importance
is of order $T_{\mathrm{spur}}\approx f_{\delta t} N_\mathrm{SS}
T_{\mathrm{relax}}$.

This effect has been tested in computations presented in
appendix~\ref{app:spur_rel}. Their result is that, provided the number
of super-stars is larger than a few thousand, there is no sign of
significant spurious evolution after a number of numerical steps
larger than what is required in any ``standard'' simulation. No
relaxation being simulated, these bare-bones steps only consist of
orbital displacements as described in Sect.~\ref{sec:selecRorb}.
Consequently, it appears that these radial movements do not introduce
appreciable spurious relaxation and that the orbital
sampling proceeds correctly.}

\subsection{Core collapse of an isolated single mass cluster}
\label{subsec:plummer_cc}

%---------------------------------------------------------------
% Table of core collapse times for Plummer model 
%---------------------------------------------------------------
\begin{table*}
  \caption{Various published values for the core collapse time of an
    isolated single mass Plummer cluster. Times are given in units of
    half-mass relaxation time $T_\mathrm{rel}^\mathrm{h}$ 
    \citep{Spitzer87}. For a Plummer model
    $T_\mathrm{rel}^\mathrm{h} = 0.093\,
    \tilde{\mathcal{U}}_\mathrm{t}$ 
    (Eq.~\ref{eq:time_unit_rel}). Numbers in parenthesis in the last
    column give the number of particles used. In the
    ``method'' column, FP stands for direct Fokker-Planck resolution
    and HMC for H{\'e}non-like Monte Carlo schemes.
    }
  \begin{center}
    \begin{tabular}{lll} \hline
      Reference&Numerical method&Core collapse time \\ \hline
      H{\'e}non \citeyearpar{Henon73,Henon75}&HMC&$\sim 14.0-18.3$ (1k)\\
      Spitzer \& Shull \citeyearpar{SS75a}&Princeton MC&$\sim 14.0-15.4$ (1k)\\
      Cohn \citeyearpar{Cohn79}&Anisotropic FP&15.9\\
      Marchant \& Shapiro \citeyearpar{MS80}&Cornell MC&14.7\\
      Cohn \citeyearpar{Cohn80}&Isotropic FP&15.7\\

      Stod\'{o}{\l}kiewicz \citeyearpar{Stodol82}&HMC&$\sim 15.7$ (1.2k)\\
      Takahashi \citeyearpar{Takahashi93}&Isotropic FP&15.6\\
      Giersz \& Heggie \citeyearpar{GH94b}&$N$-body&$\sim 17.4$ (2k)\\
      Takahashi \citeyearpar{Takahashi95}&Anisotropic FP&17.6\\
      Spurzem \& Aarseth \citeyearpar{SA96}&$N$-body&18.2 (2k), 18.0 (10k)\\
      Makino \citeyearpar{Makino96}&$N$-body&16.9 (8k), 18.3 (16k), 17.7
      (32k)\\
      Quinlan \citeyearpar{Quinlan96}&Isotropic FP&15.4\\
      Giersz (private communication)&HMC&$\sim 18.3$ (4k), $\sim
      17.5$ (64k), 17.4 (100k)\\
      Lee (private communication)&Isotropic FP&16.1\\
      Drukier et~al. \citeyearpar{DCLY99}&Anisotropic FP&17.8 ($N_\ast=8000$), 18.1 ($N_\ast=50\,000$)\\
      Joshi et~al. \citeyearpar{JRPZ00}&HMC&15.2 (100k)\\
      This work&HMC&17.8 (512k) 17.9 (2000k)\\
    \end{tabular}
\end{center}
  \label{table:tcc_plummer}
\end{table*}

%------------------------------------------------------------
% Figure : Code benchmarking
%------------------------------------------------------------

\begin{figure}
  \resizebox{\hsize}{!}{
    \includegraphics{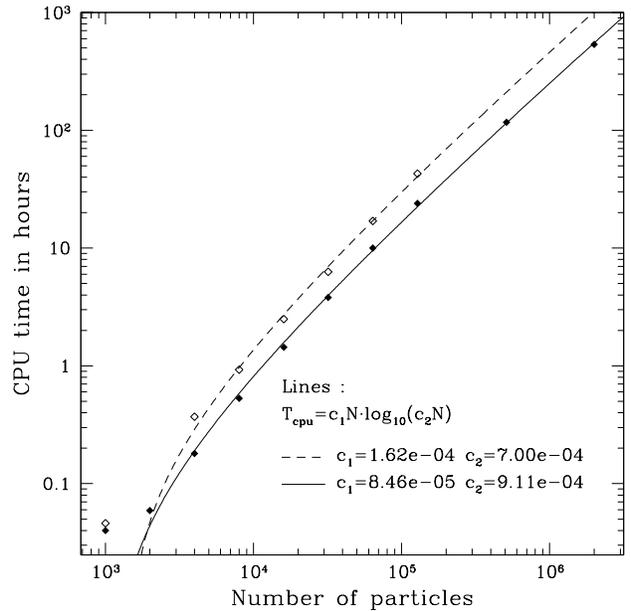}% lnN_lnTcpu.eps --> MS1125f4.eps
    }
  \caption{ Code benchmarking. CPU time to deep core collapse
    ($\Phi(0) = -10$) is shown as a function of particle (super-star)
    number for simulations of single-mass isolated Plummer clusters.
    Two sets of simulations were run with a different resolution for
    the radial mesh used to evaluate the density and the time steps.
    Open and black diamonds come from runs with 25 and 100 super-stars
    per cell, respectively. The factor 2 difference in
    $T_\mathrm{CPU}$ between both series probably originates in the
    fact that, due to Eq.~\ref{eq:delta_t_bar}, the mean time step is
    sensitive to cells with exceptionally low values of
    $T_\mathrm{rel}$ (and/or $T_\mathrm{coll}$ if collisions are
    present). Such out-lying values are due to the noise in the
    grid-evaluated $T_\mathrm{rel}$ and are smoothed out when averages
    are computed with higher number of super-stars. The lines are $
    T_\mathrm{CPU}=c_1 N_\mathrm{SS} \log_{10}(c_2 N_\mathrm{SS})$
    relations computed from least square adjustments on points for
    $N_\mathrm{SS} \geq 2000$. CPU times for simulations with
    $N_\mathrm{SS} < 2000$ are probably dominated by input-output and
    other system operations rather than by the MC algorithm itself.}
  \label{fig:plum_bench}
\end{figure}

%------------------------------------------------------------
% Figure : Evolution of Lagrangian radii in 512k plummer simul.
%------------------------------------------------------------

\begin{figure}
  \resizebox{\hsize}{!}{
    \includegraphics{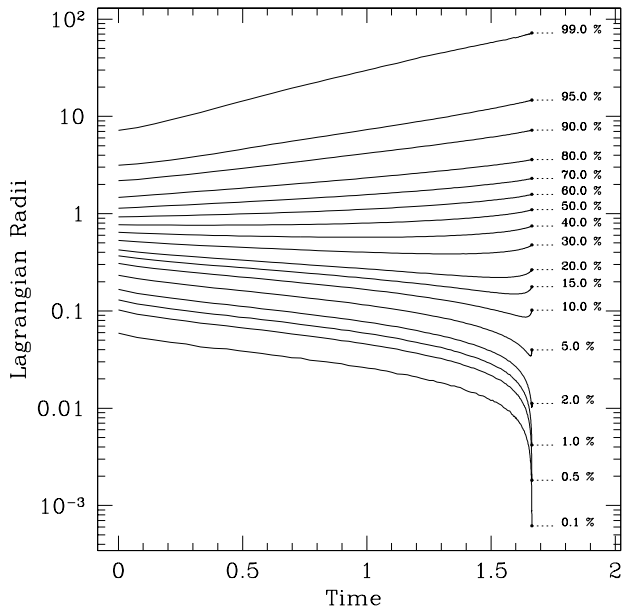}% ray_lag_PC_700.eps --> MS1125f5.eps
    }
  \caption{
    Evolution of Lagrangian radii in a core collapse simulation of an
    isolated single mass Plummer model consisting of $2\times 10^6$
    super-stars. Each curve depicts the radius of a sphere that contains
    the fraction of the mass indicated on the label at the right end.
    These fractions are given with respect to the remaining cluster's
    mass which progressively decreases due to evaporation of stars.
    ``N-body units'', $\mathcal{U}_\mathrm{l}$ and
    $\tilde{\mathcal{U}}_\mathrm{t}$ are used (see
    Sect.~\ref{sec:units}). }
  \label{fig:plum_lag_radii}
\end{figure}

%------------------------------------------------------------
% Figure : Comparison of 512k plummer with Giersz
%------------------------------------------------------------

\begin{figure}
  \resizebox{\hsize}{!}{
    \includegraphics{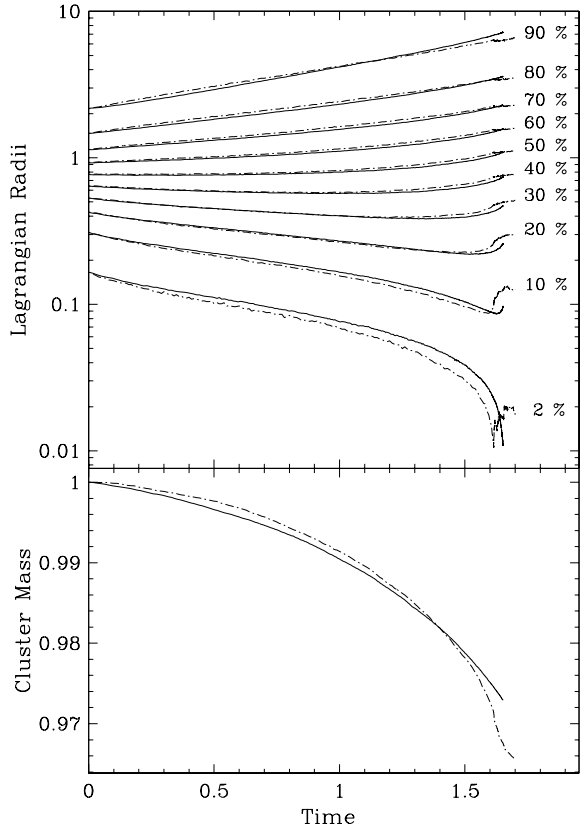}% comp_giersz_NB_PC_229.eps --> MS1125f6.eps
    }
  \caption{
    Comparison of our collapse simulation (solid lines) with a
    computation by Mirek Giersz (dash-dotted lines). Our simulation is
    the same as in Fig.~\ref{fig:plum_lag_radii}. Giersz used 100\,000
    super-stars. The top panel shows the Lagrangian radii. The bottom
    panel depicts the total mass. Contrary to ours, Giersz's code is
     able to go past core collapse by simulating the formation of
    3-body binaries and their giving up energy to other stars.}
  \label{fig:plum_comp_giersz}
\end{figure}

%------------------------------------------------------------
% Figure : Density evolution for 512k plummer
%------------------------------------------------------------

\begin{figure*}
  \centering
  \resizebox{16cm}{!}{
    \includegraphics{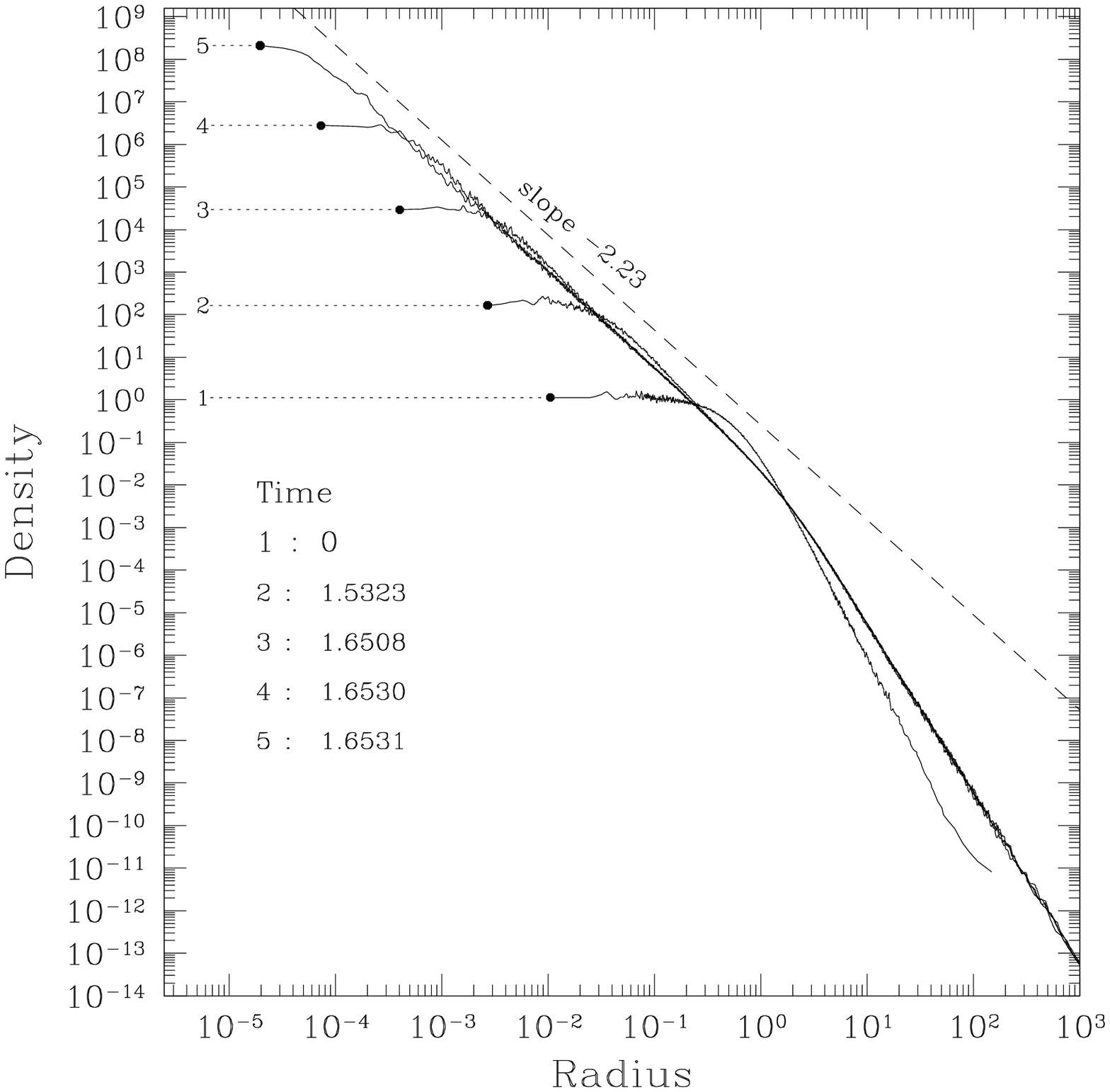}% Evol_Dens_PC_229.eps --> MS1125f7.eps
  }
  %\hfill
  %\parbox[b]{55mm}{
  \caption{
    Evolution of the density profile during core collapse for the same
    simulation as
    Figs.~\protect\ref{fig:plum_lag_radii}~and~\protect\ref{fig:plum_comp_giersz}.
    The dashed line shows the slope of the power law corresponding to
    the self-similar deep collapse solution of the Fokker-Planck
    equation \protect\citep{HS88}.}
  \label{fig:plum_dens_evol}
  %}
\end{figure*}

The next-to-simplest step was to plug in 2-body relaxation and to find
out whether we could reproduce the well studied evolution of an
isolated star cluster in which all stars have the same mass. We
chose a Plummer model because it has been extensively used in the
literature. Previous results for the collapse time are reviewed in
Table~\ref{table:tcc_plummer}. We refer to these many references for a
description and explanation of the physics of core collapse and
concentrate on some diagrams that describe our simulation of this
system and allow comparisons with others. 

The results shown here are taken from calculations with $512\,000$ and
$2\times 10^6$ super-stars which took slightly more than 100 and 500
CPU hours, respectively, to complete on a PC with a 400\,MHz
Pentium~II processor. Benchmarking of the code is presented in
Fig.~\ref{fig:plum_bench} where we plot the CPU time required to
attain a value of $\Phi(0)=-10$ for the central potential as a
function of the number of super-stars used in the calculation. As most
computing time ($T_\mathrm{CPU}$) is spend in binary tree traversals
with a number of operations that scales logarithmically, we fitted
this data with a relation
\begin{equation}
  T_\mathrm{CPU}=c_1 N_\mathrm{SS} \log_{10}(c_2 N_\mathrm{SS})
\end{equation}
with constant $c_1$ and $c_2$. This is to be contrasted with direct
$N$-body integration which, in its simplest form, requires
$\mathcal{O}(N^3)$ operations per relaxation time. 

In Fig.~\ref{fig:plum_lag_radii}, we present the evolution of
Lagrangian radii, i.e., radii of spheres that contain a given fraction
(0.1\,\% to 99\,\%) of the cluster's mass. In
Fig.~\ref{fig:plum_comp_giersz}, a subset of these radii are used in a
comparison with a simulation by Mirek Giersz (private communication).
We also plot the evolution of the decreasing total mass. Clearly, the
agreement is quite satisfactory. The most obvious difference lies in
our run leading to a somewhat slower evolution, which translates into a
core collapse time $T_\mathrm{cc}$ larger by $2-3\,\mathrm{\%}$. Given
the considerable dispersion present in the literature for the value of
$T_\mathrm{cc}$, we judge this discrepancy to be only of minor
importance.  First, due to the stochastic nature of Monte Carlo
simulations, various runs with the same code but using different
random sequences yield results that differ slightly from each other.
\modif[18]{This effect is probably of minor importance for particle
  numbers as high as $2\times 10^6$ but may affect Giersz's
  data.\footnote{To check this, we performed 5 simulations with $10^5$
    particles, using different random sequences. We got a dispersion
    of only 1\% for the value of $T_\mathrm{cc}$.}} More importantly,
we stress that although it also stems from H{\'e}non's scheme,
Giersz's code inherited the deep modifications proposed by
Stod\'{o}{\l}kiewicz and is actually very different from our
implementation. Also, as Giersz thoroughly and successfully
tested his program against $N$-body data, this comparison is very
valuable in assessing the quality of our own code.

%------------------------------------------------------------
% Figure : Plummer logarithmic density gradient
%------------------------------------------------------------

\begin{figure}
  \resizebox{\hsize}{!}{
    \includegraphics[bb=16 144 600 721,clip]{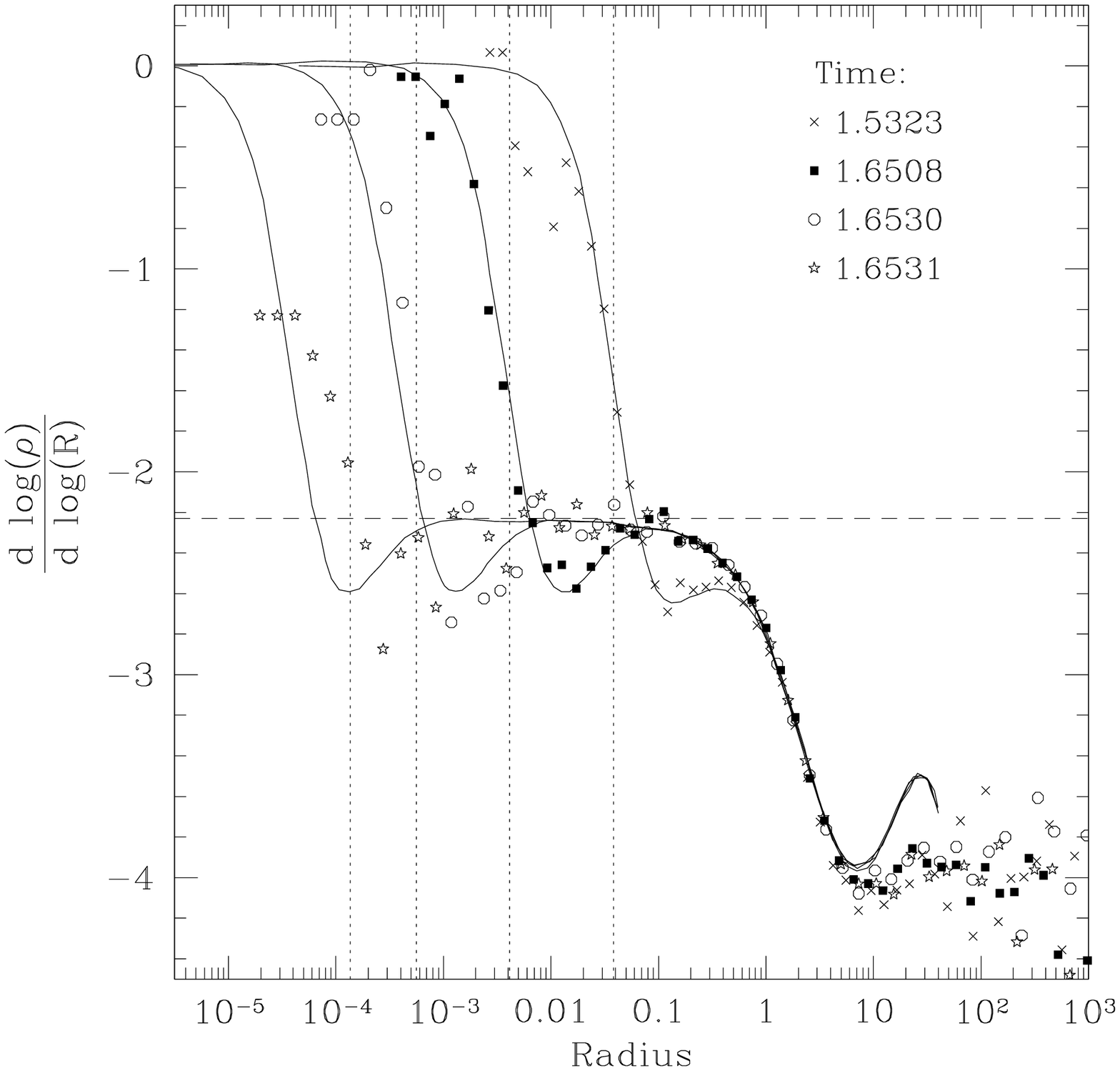}% evol_log_dens_grad_PC_229.eps --> MS1125f8.eps
    }
  \caption{
    Logarithmic density gradient for successive stages of the core
    collapse of the Plummer model. Points represent our data for the
    same times as in Fig.~\protect\ref{fig:plum_dens_evol} (except the initial
    state which is not represented here). Curves are from a simulation
    by \protect\citet{Takahashi95} for stages in core collapse with
    about the same central density increase. The leftmost curve
    corresponds to a collapse phase slightly deeper than attained by
    the Monte Carlo simulation. Dotted vertical lines show the
    decreasing values of the core radius $R_\mathrm{c} =
    \sqrt{3\langle v^2(0) \rangle / (4\pi \rho(0))}$.}
  \label{fig:plum_log_dens_grad}
\end{figure}

%------------------------------------------------------------
% Figure : Plummer Central density and dispersion evolution
%------------------------------------------------------------

\begin{figure}
  \begin{center}
    \resizebox{0.86\hsize}{!}{%
      \includegraphics{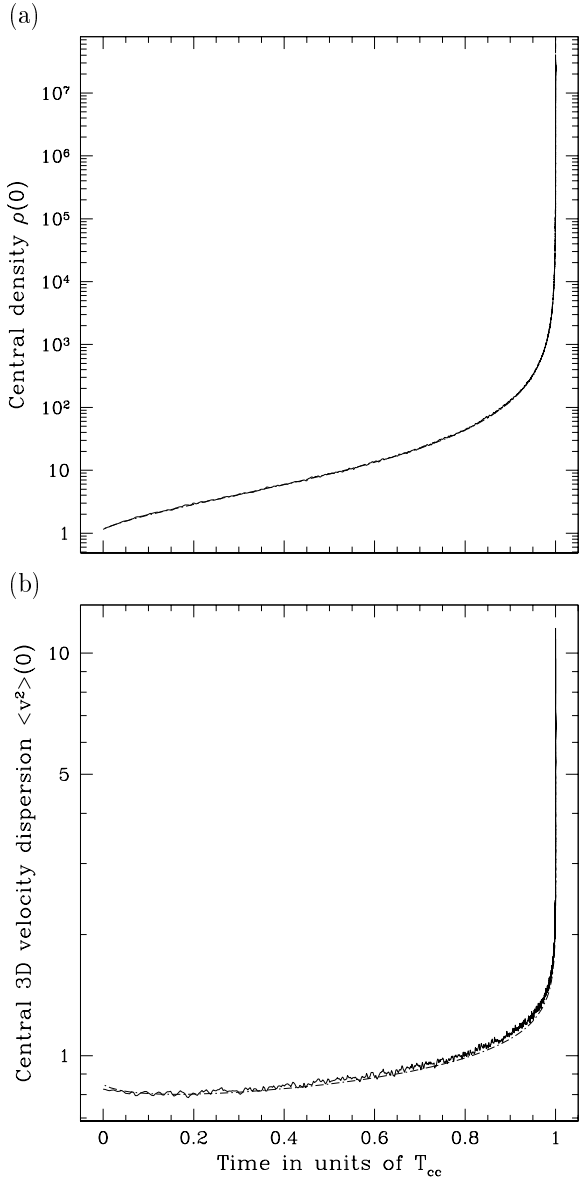}% Plummer_Comp_Lee.eps --> MS1125f9.eps
      }
  \end{center}
  \caption{
    {\bf a)} Evolution of the central density during the core collapse
    of the Plummer Model. On this diagram, the line for our simulation
    ($2\times 10^6$ particles) cannot be distinguished from an
    isotropic Fokker-Planck model by Lee (private communication)!
    \modif[25]{The time scales of both simulations have been scaled by
      their respective core-collapse times.} A slight smoothing has
    been applied to our data. {\bf b)} Same as panel a), but for the
    3D central velocity dispersion. The solid curve shows our
    simulation, the dash-dotted line is the model by Lee. }
  \label{fig:plum_ctr_dens_disp}
\end{figure}

%------------------------------------------------------------
% Figure : Plummer Anisotropy evolution
%------------------------------------------------------------

\begin{figure}
  \resizebox{\hsize}{!}{
    \includegraphics{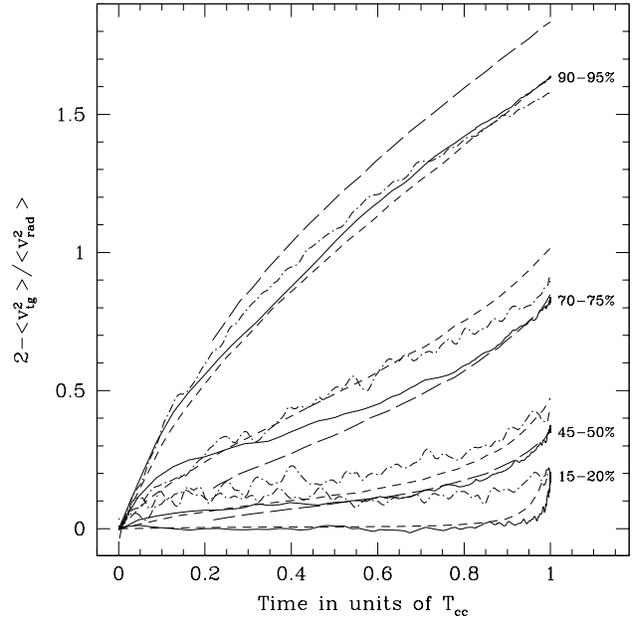}% ani_lag_PC_700.eps --> MS1125f10.eps
    }
  \caption{
    Evolution of the average anisotropy parameter within shells
    bracketed by Lagrangian spheres containing 15 to 20\%, 45 to 50\%,
    70 to 75\% and 90 to 95\% of the cluster mass. Solid lines show
    the data from our simulation with $2\times 10^6$ particles. Short
    dash curves are results from \citet{Takahashi96} for the
    anisotropy at 20\%, 50\%, 75\% and 90\% Lagrangian radii. Long
    dash curves are taken from \citet{DCLY99} for 50\%, 70\% and 90\%
    radii.  \modif[24]{Dot-dashed lines are the (smoothed) results of a $10^5$
    particle simulation by \citet{Giersz98} for 14--20\%, 44-50\%,
    70--76\% and 90--96\% shells. The time scales have been scaled by
    their respective core-collapse times. A slight smoothing has been
    applied to our and Giersz's data.}}
  \label{fig:plum_ani}
\end{figure}

Figure~\ref{fig:plum_dens_evol} is a more direct representation of the
same information. It shows the density profile $\rho(R)$ at successive
evolution phases, deeper and deeper in the collapse. According to
semi-analytical \citep[ for instance]{LBE80,HS88} and numerical
\citep[ amongst others]{Cohn80,LS91,Takahashi95,JRPZ00} computations,
the central parts of the cluster evolve self-similarly during the late
phase of core collapse according to a power-law density profile $\rho
\propto R^{-\xi}$ with $\xi \simeq 2.2$, that extends inwards. To
illustrate and confirm this behavior, in
Fig.~\ref{fig:plum_log_dens_grad}, we plot $\mathrm{d}
\ln(\rho)/\mathrm{d} \ln(R)$ versus $\ln(R)$ and compare with curves
obtained by \citet{Takahashi95} with an anisotropic Fokker-Planck
code.  Although the noisy aspect of our Monte Carlo data expectedly
contrast with the smooth curves from Takahashi's finite-difference
code, the agreement is clear. The progressive development of a
$R^{-2.2}$ in our simulation is thus well established.

As further evidence for the good performance of our code, in
Fig.~\ref{fig:plum_ctr_dens_disp}, we follow
the increase of central density $\rho(0)$ and central 3D velocity
dispersion $\langle v^2(0) \rangle$ in our model and compare them with
data from an isotropic Fokker-Planck computation by H.M.~Lee (private
communication). The collapse time of the two simulations being quite
different ($T_\mathrm{cc}=17.9$ and $16.1\,
\tilde{\mathcal{U}}_\mathrm{t}$, respectively), we
rescaled the time scale by $T_\mathrm{cc}^{-1}$ in these diagrams to
get more meaningful comparisons. Here again, the level of agreement is
more than satisfactory. Incidentally, we note that a $\langle v^2(0)
\rangle \propto \rho(0)^\zeta$ relation is quickly established with
$\zeta \simeq 0.10$ as previously noted by many authors
\citep{Cohn79,Cohn80,MS80,Takahashi95}.

Finally, in Fig.~\ref{fig:plum_ani}, we investigate the evolution of
anisotropy, measured by the usual parameter $A=2-\langle
v_\mathrm{tg}^2 \rangle / \langle v_\mathrm{rad}^2 \rangle$ where
$v_\mathrm{tg}$ and $v_\mathrm{rad}$ are the tangential and radial
components of the stellar velocity. The development of a radial
anisotropy at every Lagrangian radius is clearly visible. Our curves
are very similar to those obtained by \citet{Takahashi96} and
\citet{DCLY99} using anisotropic FP codes. Globally, although the
agreement is not as close as in previous diagrams, this relative
mismatch is weakened when examined in the light of the differences
between both FP simulations. It thus seems reasonable to think that
these differences could be due to the further simplifying assumptions
required in the derivation of direct anisotropic FP schemes.
\modif[24]{We also include the Monte Carlo simulation by
  \citet{Giersz98} in the comparison. Due to the relatively low number
  of particles used by this author ($10^5$), this data contains large
  statistical fluctuations. It is nonetheless clear that it shows
  significantly more radial anisotropy in the central regions at early
  times, compared to the other simulations. The reason for this
  difference is unknown to us but may be due to the use of radial
  ``super-zones'' by Giersz to define block time-steps. In his scheme,
  super-stars from an inner super-zone are not allowed to skip to an
  outer zone, unless both zones happen to be synchronized at the end
  of their respective time-steps. This clearly could lead to some
  ``particle restraining'' in the central parts but it is unclear why
  this effect would appear in the anisotropy profiles but not in the
  density data.}  Comparison with data from an $N$-body code would
allow us to settle these questions.  Unfortunately, present-day
accessible $N$ values (a few $10^4$) yield anisotropy curves still too
noisy to be of real use \citep[ for instance]{SA96}.

To summarize this subsection, we can safely conclude that, when
applied to the highly idealized relaxation-driven evolution of a
single-mass cluster, our Monte Carlo implementation produces results
in very nice agreement with many other modern numerical methods and
theoretical predictions. To step closer to physical realism, we now
present the results for multi-mass models.

\subsection{Evolution of clusters with two mass components}

\begin{figure}
  \resizebox{\hsize}{!}{%
    \includegraphics{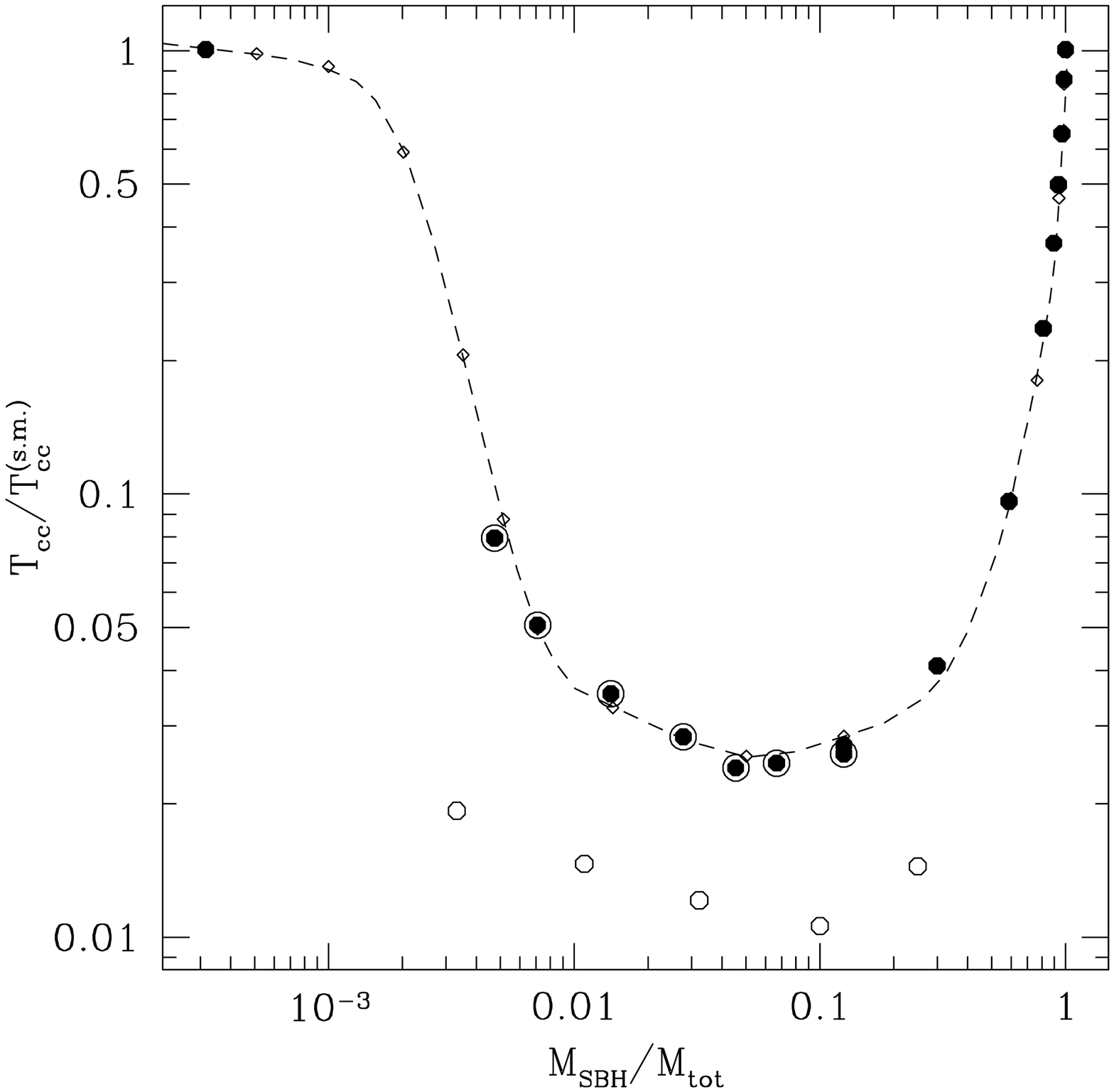}}% comp_Tcc_Lee95.eps --> MS1125f11.eps
  \caption{
    Collapse times for cluster models consisting of 2~components: MS
    stars with mass $m_1$ and stellar BHs with mass $m_2$. The initial
    clusters are Plummer models without segregation. The collapse time
    $T_{\mathrm{cc}}$ is normalized by the collapse time for a
    single-mass cluster, $T_{\mathrm{cc}}^{\mathrm{(s.m.)}}$.
    Collapse times are plotted as a function of the mass fraction of
    SBHs in the cluster.  Black dots show our MC simulations for
    $m_1=0.7\,M_{\sun}$ and $m_2=10\,M_{\sun}$. The leftmost point
    actually comes from a simulation with
    $M_{\mathrm{SBH}}/M_{\mathrm{tot}}=0$. The circled dots are for
    simulations with 512\,000 super-stars; 64\,000 super-stars have been
    used for other simulations. The open diamonds connected by the
    dashed line are data from 1-D Fokker-Planck simulations by
    \protect\citet{Lee95}. The open circles show our results for
    $m=0.3\,M_{\sun}$ and $M=10\,M_{\sun}$ with 512\,000 super-stars. }
  \label{fig:comp_Tcc_Lee95}
\end{figure}
 
\begin{figure}
  \begin{center}
    \resizebox{0.86\hsize}{!}{%
      \includegraphics{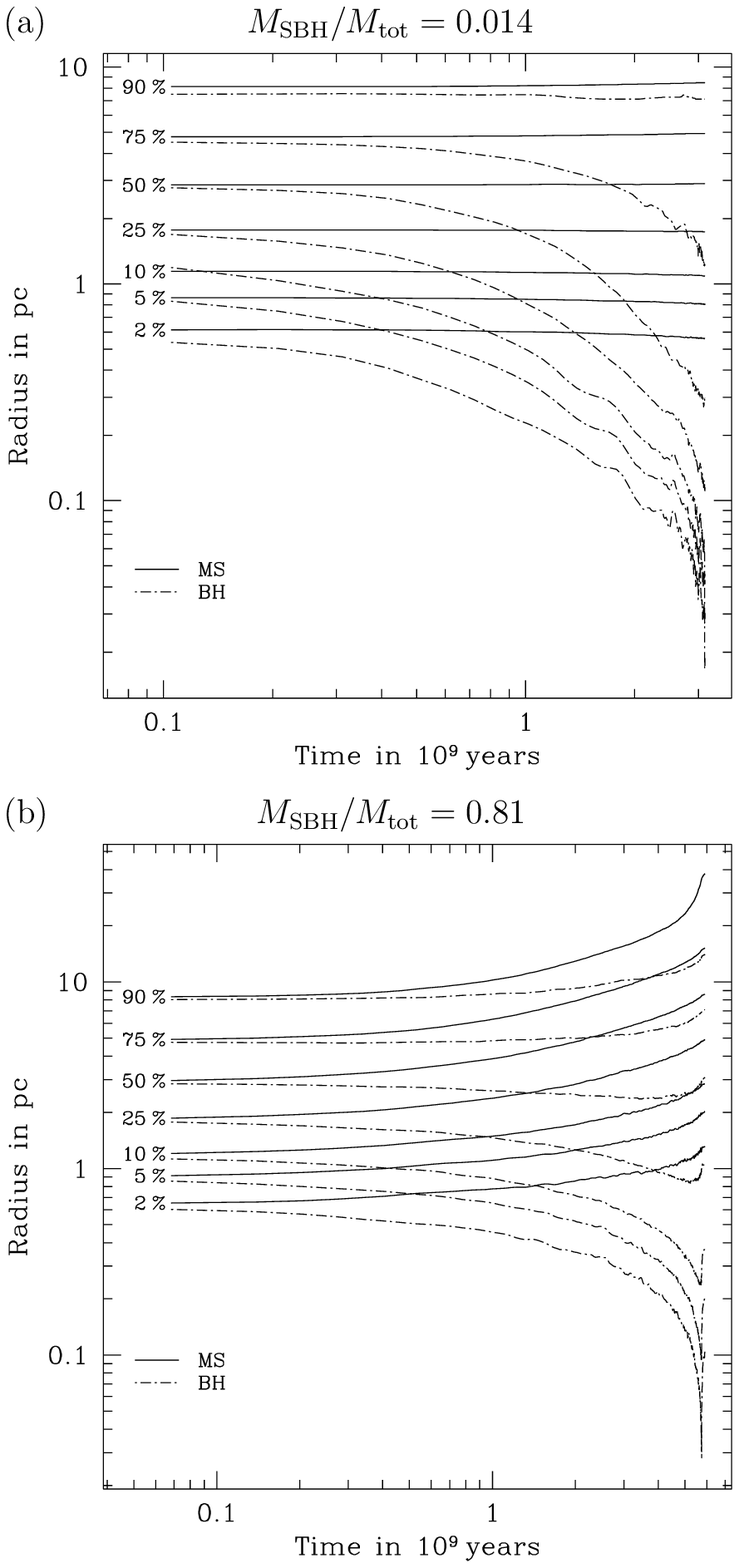}}% exples_2comp_Lee.eps --> MS1125f12.eps
  \end{center}
  \caption{
    Evolution of the Lagrangian radii of the components for two models
    of Fig.~\protect\ref{fig:comp_Tcc_Lee95}. The solid and dashed
    lines are for stars with $m_1=0.7\,M_{\sun}$ (``main sequence'': MS) and
    $m_2=10\,M_{\sun}$ (``stellar black holes'': BH), respectively.}
  \label{fig:exples_2comp}
\end{figure}      
                        
Cluster models with stars of identical masses fall short of any
realistic description of real clusters. Indeed, it has long been known
that the evolution is profoundly affected by a stellar mass spectrum
\citep[ for instance]{IW84,IS85,MC88,Takahashi97}.  If the cluster is
initiated with the same spatial and velocity distributions for the
various stellar masses, 2-body gravitational encounters will attempt
to enforce equipartition of kinetic energy between stars of different
masses, hence causing heavy stars to segregate towards the centre.
\modif[26]{Depending on the relative masses and numbers of heavy and light stars,
a temporary equipartition may be (nearly) attained
\citep{IW84,WJR00}. The latter case corresponds to the segregation (or
``stratification'') instability \citep{Spitzer87} but even when it
does not occur, a central sub-cluster containing massive stars forms
and collapses quickly because its relaxation time is much lower than
the overall value.} For a two component model, with stellar masses
$m_1$ and $m_2$ ($m_2>m_1$), the time scale for equipartition and
induced segregation can be as low as $T_\mathrm{eq} \simeq
m_1/m_2\cdot T_\mathrm{rel}$ \citep{Spitzer69}.  As a consequence, the
structure of multi-mass clusters evolves much faster than single mass
models.

As a first validation of our code in the multi-mass regime, we
simulated clusters with two mass components. Such models have been
used by \citet{Lee95} to study the fate of stellar black holes (SBHs)
in galactic nuclei. He assumed a simple stellar population with all
main sequence stars with mass $m_1=0.7\,M_{\sun}$ and a given fraction
of $m_2=10\,M_{\sun}$ SBHs. For a board range in the total mass
fraction of SBHs, his 1-D Fokker-Planck simulations show a collapse
time for the SBH sub-system that is reduced by factors of tens
compared to the single-mass case. As shown in
Fig.~\ref{fig:comp_Tcc_Lee95}, we successfully reproduce his results
for various mass fraction of SBHs.  Note that the MC method is unable
to simulate reliably clusters containing a very small SBH fraction if
the number of super-stars has to be much lower than the number of
simulated stars. \modif[28]{For instance, if we wanted to simulate a
  cluster with $M_{\mathrm{SBH}}/M_{\mathrm{tot}}=10^{-3}$ where
  $M_{\mathrm{SBH}}$ is the total mass of SBHs, with $5\times 10^5$
  super-stars, only 35 of them would represent SBHs, a number clearly
  too small to resolve the collapse of the central sub-cluster of
  SBHs.} To illustrate the process of mass segregation, in
Fig.~\ref{fig:exples_2comp} we plot the evolution of the Lagrangian
radii of both mass components for two of these cluster models.

\subsection{Evolution of a tidally truncated multi-mass cluster}

%------------------------------------------------------------
% Figure : Collaborative experiment : Lagrangian radii
%          comparison with Giersz.
%------------------------------------------------------------

\begin{figure}
  \resizebox{\hsize}{!}{ \includegraphics{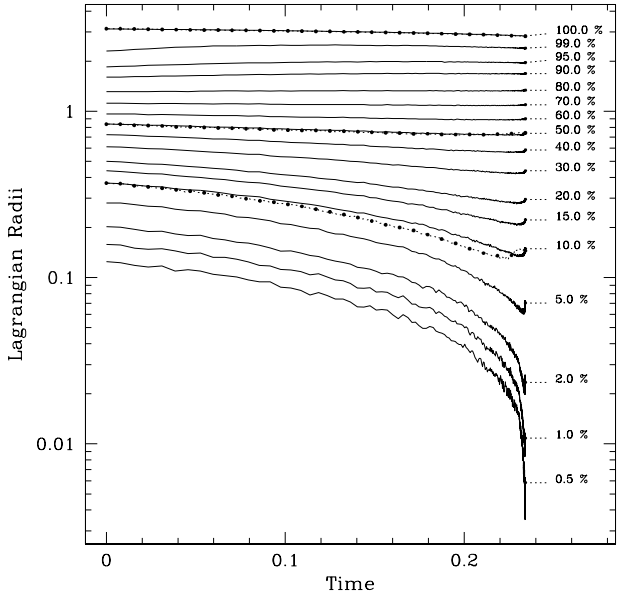} }% ray_lag_CE.eps --> MS1125f13.eps
  \caption{
    Evolution of the Lagrangian radii for a cluster with initial
    conditions according to Heggie's ``collaborative
    experiment''. Solid lines are the result of our simulation with
    256\,000 super-stars. Dashed lines with black dots are from a
    computation by Mirek Giersz (10, 50 and 100\% radii). The tidal
    radius is label as ``100\%''. ``$N$-body'' units are used. Unit of
    time: $5.70\times10^{10}$\,yrs. Unit of length: $9.55\,
    \mathrm{pc}$.  }
  \label{fig:CE_ray_lag}
\end{figure}

%------------------------------------------------------------
% Figure : Collaborative experiment : mass segregation
%          comparison with Giersz.
%------------------------------------------------------------

\begin{figure}
  \resizebox{\hsize}{!}{ \includegraphics{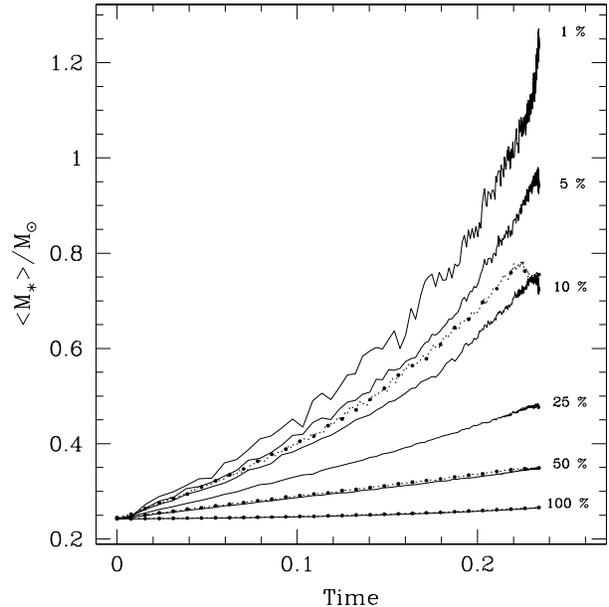} }% segr_CE.eps --> MS1125f14.eps
  \caption{
    Mass segregation diagram for a cluster with initial
    conditions according to Douglas Heggie's ``collaborative
    experiment''. Each curve show the evolution of the average mass of 
    stars in a sphere that contains a given fraction of the total
    cluster's mass, as indicated by labels on the right.
    Solid lines are the result of our simulation with
    256\,000 super-stars. Dashed lines with black dots are from a
    computation by Mirek Giersz (10, 50 and 100\% radii). The unit of
    time is $5.70\times10^{10}$\,yrs.
    }
  \label{fig:CE_avrg_mass}
\end{figure}

To study the evolution of a cluster with a continuous mass function,
we simulated a model with initial conditions set according to Heggie's
``collaborative experiment''\footnote{All experiment data is available
  at \texttt{http:// www.maths.ed.ac.uk/$\sim$douglas/experiment.html}}
\citep{HGST99}. This is a King model with $W_0=-\Phi(o)/\sigma^2=3$
\citep[ p.232]{BT87}. A spherical tidal truncation is imposed at
$R_\mathrm{tid}=30\,\mathrm{pc}$. The mass spectrum is:
\[
\frac{\mathrm{d}N}{\mathrm{d}M_\ast} \propto M_\ast^{-2.35}
\mbox{~~for~} 0.1\,M_{\sun} \leq M_\ast \leq 1.5\,M_{\sun}
\] 
and the total mass is $6\times10^4 M_{\sun}$. Hence, the number of
stars is $N_\ast = 2.474\times10^5$. There is no initial mass
segregation and no primordial binaries. According to the rules of the
experiment, no stellar evolution has to be simulated but the heating
effect of binaries \modif[30]{could be incorporated to simulate the
  post-collapse evolution up to complete evaporation of the cluster.}

Our code lacks the ability to simulate the formation of binaries and
their heating effect. However, as explained in
Sect.~\ref{sec:binaries}, these processes do not switch on before the
core has collapsed down to a few tens of stars. As a consequence, we
should be able to tackle the evolution of this system up to deep core
collapse.

Many researchers, using a variety of simulation methods, from gas
models to $N$-body codes, have taken part in the ``collaborative
experiment''. Their results show a very important dispersion. For
instance, the obtained core-collapse times range from 9 to more than
14\,Gyrs while the values for cluster's mass at this time lie between
$2.2\times 10^4$ and $4.75\times 10^4\,M_{\sun}$! In our simulations
with 16\,000 to 256\,000 super-stars, we find a collapse time of 12.5 to
13.4\,Gyrs with a remaining mass varying from $4.63\times 10^4$ to
$4.37\times 10^4\,M_{\sun}$. 

Factor 2 discrepancies can even been found amongst simulations using
the same scheme, e.g. $N$-body codes.  There is a clear tendency for
$N$-body to yield values of $T_\mathrm{cc}$ shorter than those
produced by other, statistical, methods. Another perplexing fact is
that the results of $N$-body simulations do not converge to those of
statistical methods as $N$ increases, contrary to naive expectations.
\modif[31]{The cause of these unexpected results has been traced by \citet{FH99}
to two combining facts. First, for the realistic, non-spherical, tidal
potential used in those simulations, stars with energies above the
escape energy can stay in the cluster for many dynamical times before
they actually leave it. Second, the way the models where initiated led
to clusters containing, from the beginning, a large fraction of such
``potential escapers'', instead of being in equilibrium in the tidal
field.}

Given this confusing picture, it seems more sensible to compare our
results to those produced by a similar computational approach. Mirek
Giersz applied his Monte Carlo code \citep{Giersz98,Giersz00b} to this
system. In Fig.~\ref{fig:CE_ray_lag}, we show the evolution of
Lagrangian radii for his simulation (up to binary-induced rebound) and
ours. Similarly, Fig.~\ref{fig:CE_avrg_mass} compares the evolution of
the average mass of stars. This latter diagram clearly shows how
strong mass segregation effects are in multi-mass clusters. The
relatively good agreement to be read from these figures supports our
code's ability to handle star clusters with a mass spectrum.

\section{Conclusions}
\label{sec:concl}

\subsection{Summary}

In this paper, we have presented a new stellar dynamics code we have
recently developed. It can be seen as a Monte Carlo resolution of the
Fokker-Planck equation for a spherical star cluster. Although stemming
from a scheme invented by H\'enon in the early 70's, it was deemed
optimal for our planned study of the long-term evolution of dense
galactic nuclei hosting massive black holes. The main advantages of
this kind of approach are a high computational efficiency (compared to
$N$-body codes), on the one hand, and the ability to incorporate many
physical effects with a high level of realism (compared to direct
Fokker-Planck resolution or to gas methods), on the other hand.  These
features explain the recent revival of H\'enon's approach in the realm
of globular cluster dynamics by \citet{Giersz98} and \citet{JRPZ00}.
To the best of our knowledge, however, we are the first to apply it to
galactic nuclei (see \citealt{FB01d,FB01b}).

The version of the code presented here only includes 2-body relaxation.
Spherically symmetric self-gravitation is computed exactly. Arbitrary
mass spectrum and velocity distribution, isotropic or not, can be
handled without introducing any extra computational burden. The test
computations we carried out allow us to be confident in the way our
code simulates the evolution of spherical star clusters over the long
term, as driven by relaxation.

The computational speed of our code is highly satisfying. The
evolution of a single-mass globular cluster with 512\,000 super-stars
up to core collapse takes about 5~CPU~days on a 400\,MHz Pentium-II
processor. This can be compared with the three months of computation
required by \citet{Makino96} to integrate a cluster with 32\,000
super-stars on a GRAPE-4 computer specially designed to compute forces
in an $N$-body algorithm. However, the significance of such a
comparison is somewhat blurred by the fact that Makino integrated the
system past core collapse and that the hardware in use is so
different.  Nevertheless, the speed superiority of our Monte Carlo
scheme over $N$-body really lies in a CPU time scaling as
$N\cdot\log(cN)$ instead of $N^{2-3}$ (Makino reports $T_\mathrm{CPU}
\propto N^{2.3}$.). Furthermore, Monte Carlo simulations do not have
to resolve orbital time-scales; their time step is a fraction of the
relaxation time which is of an order $10^5$ times larger for a
million-star self-gravitating cluster. This ensures that Monte Carlo
simulations will remain competitive in the next few years, even after
the advent of special-purpose $N$-body computers with highly increased
performances like the GRAPE-6 system \citep{Makino98,Makino00}. Monte
Carlo codes like ours are bound to become the tool of choice to
explore the dynamics of star clusters by allowing investigators to run
many simulations with a variety of initial conditions and physical
processes. Run-of-the-mill personal computers are sufficient to get
quick results without sacrificing too much of the physical realism.

\subsection{Other physical ingredients and future developments}
\label{subsec:future}

2-body relaxation is only one amongst the many physical processes that
are thought to contribute to the long term evolution of dense galactic
nuclei or are of high interest of themselves even without a global
impact on the cluster. Here, we list the most important of them and
comment on those not already discussed in Sect.~\ref{sec:intro}. \modif[33]{The
order in this list broadly reflects the probable order of inclusion of
these effects in our simulations.}

\begin{itemize}
  \item Stellar collisions.
  \item Tidal disruptions. 
  \item Stellar evolution.
  \item Capture of stars by the central BH through emission of
    gravitational radiation (``GR-captures''). As these events are a
    very promising source of gravitational waves for the future
    space-borne interferometer LISA, reliable predictions for their
    rate and characteristics are highly desirable even though they are
    unlikely to play a dominant role in the BH's growth
    \citep{Danzmann96,Thorne98}.
  \item Large scale gas dynamics. Gas is released by stars during
    their normal evolution (winds, SN explosions,\ldots) or as a
    consequence of collisions. Recent 2-D hydrodynamic simulations
    by \citet{WBP99} have revealed a variety of behaviours that were
    not captured by previous works \citep{Bailey80,DDC87a,DDC87b}. The
    determination of the fraction of gas accreted by the central BH
    and the fraction that escapes the galactic nucleus appears to be a
    difficult but important problem.
  \item Interplay with outer galactic structures. Contrary to globular
    clusters, the nucleus of a non-interacting galaxy is not subject
    to a strong tidal field. However, it is not an isolated cluster.
    It is embedded in a larger structure (bar, bulge, elliptical
    galaxy) whose gravitational potential is generally not spherically
    symmetric and with which it can exchange stars and gas.
  \item Interactions with binary stars. This has been discussed in
    Sect.~\ref{sec:binaries}.
  \item Other interactions between the central black hole and stars. A
    number of more or less exotic mechanisms have been proposed in the
    literature, most of them as alternate mechanisms to feed the
    central BH with stellar fuel. Amongst others, we mention tidal
    capture of stars \citep{NPP92}, their interaction with a central
    accretion disk \citep[ and others]{Rauch95,AZD96}, mass transfer
    to the BH by a close orbiting star \citep{HKLA94} and the
    influence of the UV/X-ray flux from the accreting BH on the
    structure and evolution of nearby stars (for instance X-ray
    induced stellar winds, see \citealt{VS88}).
  \item Cluster rotation. A few recent observations indicate that the
    centre-most regions in a cluster may present substantial amounts
    of rotation \citep{GPEGO00,GPOCWH00}. 
\end{itemize}

The original H\'enon's code was devised to study idealized globular
clusters whose evolution is solely driven by relaxation. Such models
are only remotely connected to galactic nuclei. Unfortunately, the
processes possibly at play in galactic nuclei are so numerous and (for
many of them) uncertain that fully consistent simulations,
incorporating all the physics, seem to be beyond reach for many years
still. Such simulations would look misleadingly realistic but yield
little insight into the importance of each individual process and how
it interplays with the others. To favor such an understanding, we
restrict our discussion, for the time being, to a few ingredients that
are deemed particularly important. We want to get a good insight into
the ``workings'' of these simplified models before we add more
complexity and uncertainties by including further physics.
\modif[33]{Some of these ingredients are very likely to play a key
  role in the evolution of the cluster: tidal disruptions, stellar
  evolution, maybe collisions\ldots Other processes, like GR-captures,
  may be too rare to have an noticeable influence on the overall
  dynamics and structure of the system but have great observational
  promise as individual events.}

In the following paper of this series \citep{FB01b}, we'll describe
how stellar collisions and tidal disruptions are treated. The next
effects to which we shall turn are stellar evolution, included in a
simplified way in the latest version of the code \citep{MaThese}, and
GR-captures, for which encouraging results have already been obtained
\citep{Freitag01}. 
%Our preliminary simulations with stellar evolution
%pointed to the necessity for a prescription to tackle to fate of the
%liberated gas, and in particular to determine 

\begin{acknowledgements}
  
  We warmly thank Mirek Giersz and Hyung Mok Lee for helpful
  discussions and their providing unpublished simulation data. Mirek
  Giersz has also been a very efficient referee whose comments and
  suggestions have helped to improve the manuscript of this paper.
  Most simulations have been realized on the GRAVITOR ``Beowulf''
  computing farm at Geneva
  Observatory\footnote{\texttt{http://obswww.unige.ch/$\sim$pfennige/gravitor/
      gravitor\_e.html}}, with invaluable help from Daniel Pfenniger.
  This work has been supported in part by the Swiss National Science
  Foundation.

\end{acknowledgements}

% APPENDICES
%%%%%%%%%%%%%

\appendix

\section{Practical computation of a super-encounter}
\label{app:SE}

\begin{figure}
  \resizebox{\hsize}{!}{
    \includegraphics{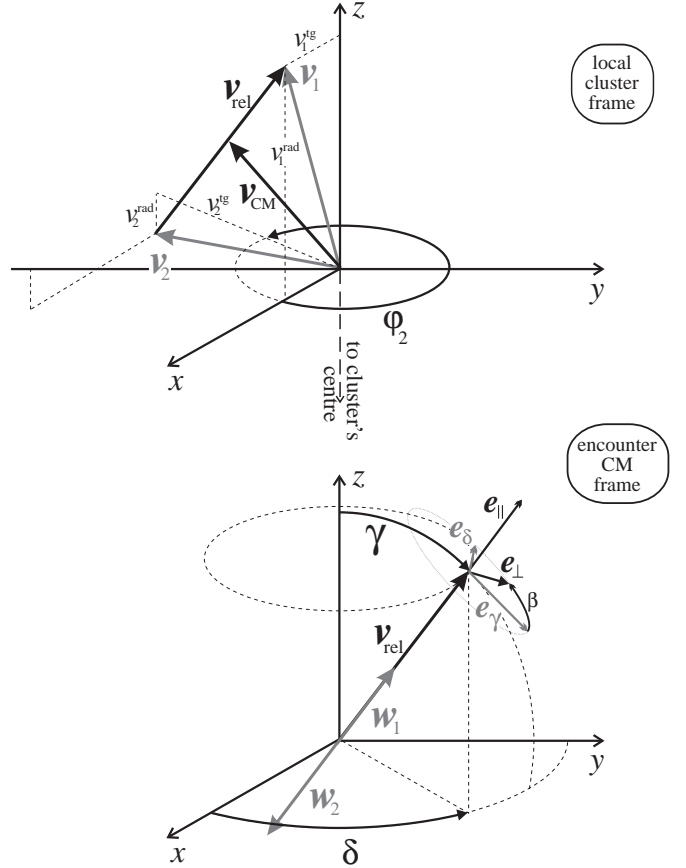}% 3d_encounter_velocities_b.eps -->  MS1125fA1.eps
    }
  \caption{The reference frames used in the computation of the 2-body
    encounter.  $\vec{v}_{1,2}$ are the velocities of the stars in the
    local cluster frame, $\vec{w}_{1,2}$, their velocities in the
    encounter reference frame,
    $\vec{v}_{\mathrm{rel}}=\vec{v}_1-\vec{v}_2$ is the relative
    velocity and $\vec{v}_{\mathrm{CM}}$ the velocity of their
    centre-of-mass (CM) in the cluster frame. See text for more
    details.}
  \label{fig:3d_encounter_velocities}
\end{figure}
 
In this appendix, we detail the vectorial operations corresponding to
steps 2, 3, and 4 of an individual encounter between two neighbouring
super-stars, as described in Sect.~\ref{sec:super_encounter}.

{\bf Step 2.} The situation is depicted on top of
Fig.~\ref{fig:3d_encounter_velocities}. A local reference frame, at
rest in the cluster, is defined with axis $\mathrm{O}z$ pointing in the
radial direction from the cluster's centre and
$\vec{v}_1\in\mathrm{O}xz$. From the specific angular momenta $J_i$, kinetic
energies $T_i$ and distances to centre $R_i$ of the super-stars, the
moduli of the radial and tangential components of the stars'
velocities are deduced:
\begin{equation}
  v_i^\mathrm{tg}=J_i/R_i\mbox{,\ \ }
  v_i^\mathrm{rad}=\sqrt{2T_i-(v_i^\mathrm{tg})^2}.
\end{equation}
We set $v_1^x=v_1^\mathrm{tg}$, $v_1^y=0$ and
$v_1^z=v_1^\mathrm{rad}$, $v_2^x = v_2^\mathrm{tg}\cos(\varphi_2)$,
$v_2^y = v_2^\mathrm{tg}\sin(\varphi_2)$ and $v_2^z = \pm
v_2^\mathrm{rad}$, with a random value for the angle $\varphi_2$ ($\in
[0,2\pi]$) and the sign of $v_2^z$. This procedure accounts for the
freedom in the relative orientation of $\vec{v}_1$ and $\vec{v}_2$,
thus ensuring a correct sampling of the encounters' incoming
velocities.

{\bf Step 3.} We now switch to the encounter CM reference frame (see
bottom panel of Fig.~\ref{fig:3d_encounter_velocities}) which has the
same axis orientation as the cluster frame but translates with
velocity $\vec{v}_\mathrm{CM}=\lambda_1\vec{v}_1+\lambda_2\vec{v}_2$
($\lambda_i=M_i/(M_1+M_2)$ where $M_{1,2}$ are the masses of the
stars). We use the notation $\vec{v}_{1,2}$ for the stars' velocities
in the cluster frame and $\vec{w}_{1,2}$ when we express them in the
encounter frame. Prime ($^\prime$) denotes quantities after the
encounter.  We build an orthonormal vector set
$\{\vec{e}_\parallel,\vec{e}_{\gamma},\vec{e}_{\delta}\}$ with
$\vec{e}_\parallel = \vec{v}_\mathrm{rel}/v_\mathrm{rel}$ and
$\vec{e}_{\gamma},\vec{e}_{\delta} \perp \vec{v}_\mathrm{rel}$. The
orientation of the orbital plane spanned by
$\{\vec{e}_\parallel,\vec{e}_\perp\}$ is set through a randomly chosen
angle $\beta$ defining $\vec{e}_\perp = \cos(\beta)\vec{e}_{\gamma} +
\sin(\beta)\vec{e}_{\delta}$. In this frame, the effect of the
gravitational encounter is simply to rotate the initial velocities (at
infinity) $\vec{w}_1=\lambda_2\vec{v}_\mathrm{rel}$ and
$\vec{w}_2=-\lambda_1\vec{v}_\mathrm{rel}$ by an angle
$\theta_\mathrm{SE}$, so:
\begin{equation}
  \begin{array}{l}
    \vec{w}_1^\prime = \lambda_2 \vec{v}_\mathrm{rel}^\prime \mbox{,\ \ }
    \vec{w}_2^\prime = -\lambda_1 \vec{v}_\mathrm{rel}^\prime \\
    \mbox{with\ \ } 
    \vec{v}_\mathrm{rel}^\prime = v_\mathrm{rel} \left(
      \cos(\theta_\mathrm{SE}) \vec{e}_\parallel +
      \sin(\theta_\mathrm{SE}) \vec{e}_\perp \right).
  \end{array}
\end{equation}

{\bf Step 4.} Finally, the relevant post-encounter properties of the
super-stars in the cluster frame are given by straightforward
formulae:
\begin{eqnarray}
  T_i^\prime &=& \frac{1}{2} (\vec{v}_i^\prime)^2, \\
  J_i^\prime &=& R_i\sqrt{(v_i^{x\prime})^2+(v_i^{y\prime})^2}
\end{eqnarray}
with $\vec{v}_i^\prime=\vec{w}_i^\prime+\vec{v}_\mathrm{CM}$.

\section{Random selection of a new orbital radius}
\label{app:placement}

\begin{figure}%[!h]
  \resizebox{\hsize}{!}{
    \includegraphics{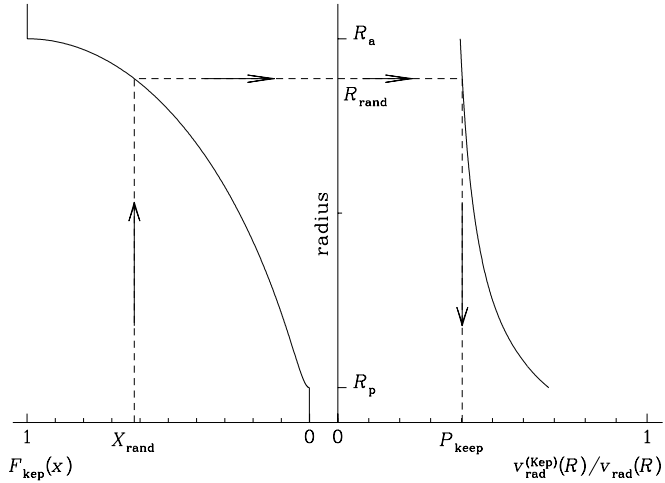}% diag_TirOrb_trans.eps --> MS1125fB1.eps
    }
  \caption{
    Diagram of the random selection of an orbital position using a
    rejection method with a Keplerian comparison function. A constant
    $P_\mathrm{selec}(R)$ is assumed.}
  \label{fig:RorbSelec}
\end{figure}

\begin{figure}%[!h]
  \resizebox{\hsize}{!}{
    \includegraphics{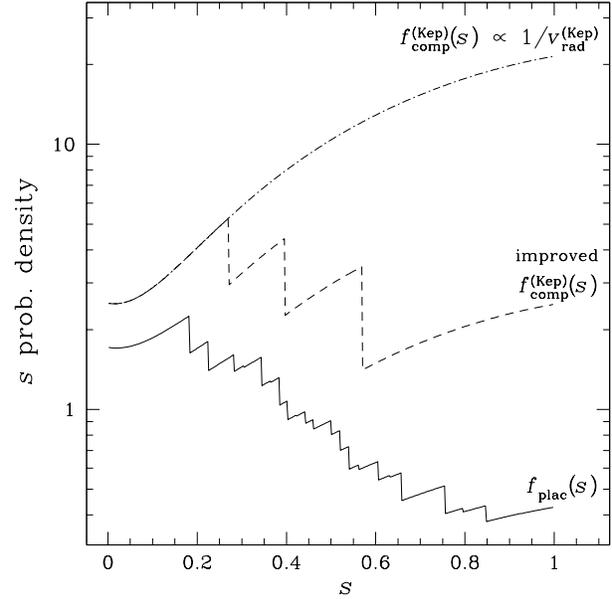}% dPds_TirOrb1.eps --> MS1125fB2.eps
    }
  \caption{
    Example of the shape of the probability density $f_\mathrm{plac}
    \propto P_\mathrm{selec}(R) / v_\mathrm{rad}(R)$ for placing a
    super-star on its orbit. H\'enon's $s$ variable has been used
    instead of $R$ to remove the divergences of $v_\mathrm{rad}^{-1}$
    at peri- and apo-centre ($R = R_\mathrm{p} +
    (R_\mathrm{a}-R_\mathrm{p})s^2(3-2s)$). $f_\mathrm{plac}$ is shown
    in solid line along with the simple Keplerian upper bound
    $P_\mathrm{selec}(R_\mathrm{p}) /
    v_\mathrm{rad}^{(\mathrm{Kep})}(R)$ (dot-dashed line) and a
    refined Keplerian upper bound (dashes) with coarse account for the
    important radial decrease of $P_\mathrm{selec}$.
    $P_\mathrm{selec}$ was computed with the second method described
    in Sect.~\protect\ref{sec:Rel_Pair_Choice} where
    $T_{\mathrm{rel}}^{-1}$ is evaluated on a coarse rank grid. This
    is responsible for the saw tooth shape of $f_\mathrm{plac}$.
    Obviously, the ``raw'' Keplerian $f_\mathrm{comp}$ would mostly
    generate $R$ values near the apo-centre very likely to be rejected
    due to the actual very low probability density for such $R$s. }
  \label{fig:dPdsRorbSelec}
\end{figure}

Here, we explain how we determine a new radius $R$ for a super-star after
it has experienced a super-encounter (see Sect.~\ref{sec:selecRorb}).

Our goal is to generate random $R$ values whose distribution density
comply with Eq.~\ref{eq:dPplacdR}. The main difficulty lies in the
fact that $f_\mathrm{plac}(R)$ is not an easily computable function.
First, $P_\mathrm{selec}$ really depends on the rank rather than on $R$ and
has generally no simple analytical expression (see
Sect.~\ref{sec:Rel_Pair_Choice}). Furthermore, $v_\mathrm{rad}(R)$ is a
function of $\Phi(R)$ which is not known analytically either.
Obtaining the value of the rank or the potential (through local $A$
and $B$ coefficients) at any given $R$ implies a tree traversal. Given
these intricacies, we don't even attempt transforming a
uniform-deviate random variable through the inverse function of the
cumulative probability and turn to a \emph{rejection} method \citep[ 
Sect.7.3]{PTVF92}. The trick is to find a \emph{comparison function}
$f_\mathrm{comp}$, a more docile probability density which can be made
a uniform upper bound to $f_\mathrm{plac}(R)$ ($f_\mathrm{comp}(R) \ge
\alpha f_\mathrm{plac}(R)\; \forall R \in [R_\mathrm{p},R_\mathrm{a}]$
for some constant~$\alpha$). We then proceed by generating a random
number $R_\mathrm{rand}$ according to $f_\mathrm{comp}$ and another,
$X_\mathrm{rand}$, with uniform deviation between 0 and 1. If
$X_\mathrm{rand} \le \alpha f_\mathrm{plac}(R_\mathrm{rand}) /
f_\mathrm{comp}(R_\mathrm{rand})$, $R_\mathrm{rand}$ is kept,
otherwise it is rejected and we try again.  The accepted
$R_\mathrm{rand}$ values follow the $f_\mathrm{plac}$ distribution. Of
course the closer the comparison function is to $f_\mathrm{plac}(R)$,
the fewer rejection steps and the more efficient is the method
(remember that a tree traversal is realised for each
$f_\mathrm{plac}(R)$ evaluation).

By construction, $P_\mathrm{selec}$ is constrained to decrease with
rank/radius, so an easy but inefficient upper bound is its value at
peri-centre. As for $v_\mathrm{rad}(R)^{-1}$, we first applied
H\'enon's variable change ($R = R_\mathrm{p} +
(R_\mathrm{a}-R_\mathrm{p})s^2(3-2s)$) to remove the divergences at
peri- and apo-centre but failed to find a safe upper bound for the
resulting probability density of $s$. We instead use the fact that a
shifted Keplerian potential $\Phi_\mathrm{Kep}(R) = C_1/R + C_2$ equal
to $\Phi$ at $R_\mathrm{p}$ and $R_\mathrm{a}$ is everywhere higher
(or equal) in between, so that
\begin{eqnarray}
  \frac{1}{v_\mathrm{rad}(R)} &\le& 
  \frac{1}{v_\mathrm{rad}^{(\mathrm{Kep})}(R)} = \nonumber \\ 
  & &
  \frac{\sqrt{R_\mathrm{p}R_\mathrm{a}}}{2J}
  \frac{R}{\sqrt{
      \left( R-R_\mathrm{p} \right)
      \left( R_\mathrm{a}-R \right) 
      }}.
\end{eqnarray}
Furthermore, the cumulative probability function for this Keplerian
bound is analytical:
\begin{eqnarray}
  F_\mathrm{Kep}(R) &\propto& \int_{R\mathrm{p}}^R 
  \frac{ r\,\mathrm{d}r }{ \sqrt{
      \left( r-R_\mathrm{p} \right)
      \left( R_\mathrm{a}-r \right) 
      } } \nonumber \\
  &=& \frac{2}{\pi} \left[ \arctan\sqrt{\frac{x}{1-x}} -
    e\sqrt{x(1-x)} \right] \\
  \lefteqn{ \mbox{with~~} x =
    \frac{R-R\mathrm{p}}{R\mathrm{a}-R\mathrm{p}}  \mbox{~~and~~}
    e=\frac{R\mathrm{a}-R\mathrm{p}}{R\mathrm{a}+R\mathrm{p}} }.&& \nonumber
\end{eqnarray}
So, to summarize, the selection of a new orbital radius proceeds as
follows (see Fig.~\ref{fig:RorbSelec}):
\begin{enumerate}
  \item Generate a random number $X_\mathrm{rand}$ with
    $[0,1]$-uniform deviate.
  \item Compute $R_\mathrm{rand} = R_\mathrm{p} +
    (R\mathrm{a}-R\mathrm{p}) F^{-1}_\mathrm{Kep}(X_\mathrm{rand})$
    using, for instance, Newton-Raphson algorithm.
  \item Keep $R=R_\mathrm{rand}$ with probability 
    \[
    P_\mathrm{keep} = \frac{
    P_\mathrm{selec}(R) }{ P_\mathrm{selec}(R_\mathrm{p}) }
  \frac{
    v_\mathrm{rad}^{(\mathrm{Kep})}(R) }{ v_\mathrm{rad}(R) }.
    \]
\end{enumerate}

This procedure actually had to be improved, for $P_\mathrm{selec}$ is
generally very rapidly decreasing with rank. As a result, if the
super-star's orbit spans a large $R$-range, a constant bound often
proves to be highly inefficient, requiring hundreds of rejection
steps.  So, when the number of unsuccessful tries reaches a limiting
value, the $[R_\mathrm{p},R_\mathrm{a}]$ interval is sliced into a few
sub-intervals and a stepped bound on $P_\mathrm{selec}$ is devised by
computing its value at the lower rank of each sub-interval. Hence a
comparison function is obtained that lies closer to $f_\mathrm{plac}$.
However, in case a large number of rejections still fails to select a
$R$ value, further successive refining is realized, using more and
more sub-intervals to get closer and closer upper bounds to
$P_\mathrm{selec}$. In practice, we use successively 5, 20 and 50
sub-intervals when the number of unsuccessful rejection cycles exceeds
10, 100 and 1000 respectively.  This modification clearly complicates
the computation of $F_\mathrm{Kep}$ and its inversion during phase 2.
It appears as a numerical overhead as it imposes a tree traversals to
determine lower rank values for each sub-interval. However, such added
intricacies allow to break off from a (nearly) never ending rejection
cycle. The improvement on the Keplerian comparison function is
depicted for a typical case in Fig.~\ref{fig:dPdsRorbSelec}. With this
method, the average number of rejection cycles to attribute a new $R$
to a super-star lies between 5 and 10 in our cluster simulations.

\section{Binary tree structure and algorithms}
\label{app:bin_tree}

\begin{figure*}
  \resizebox{12cm}{!}{
    \includegraphics{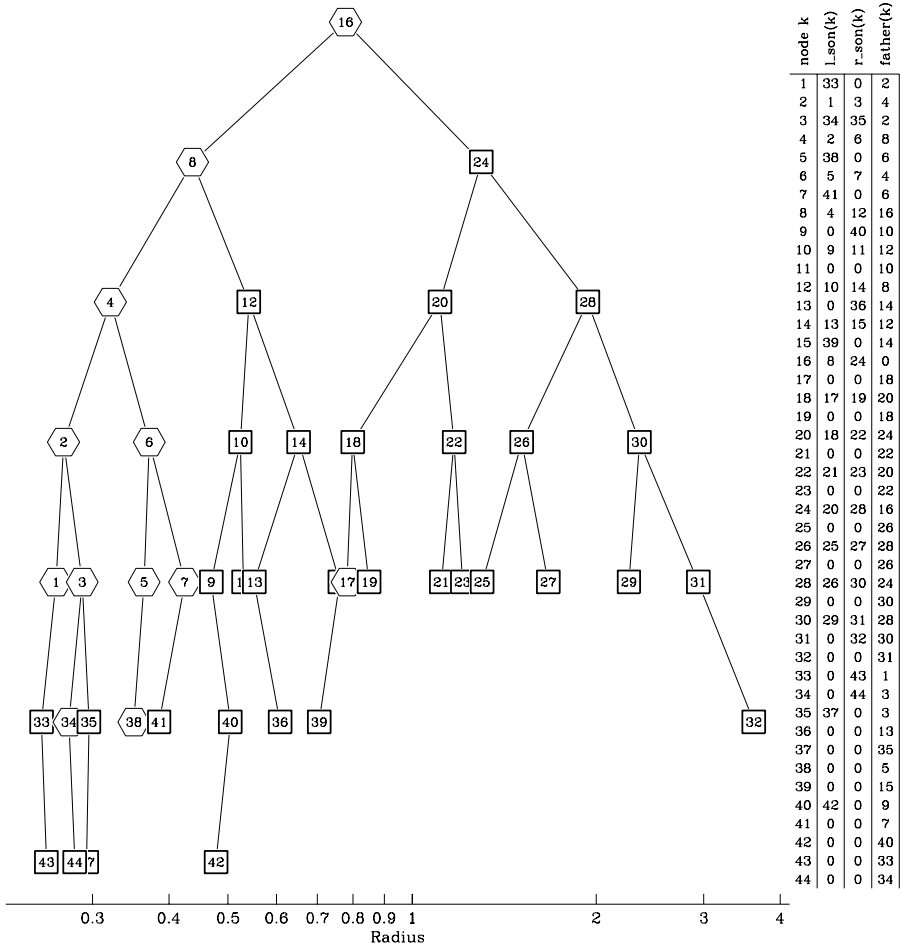}% bin_tree1.eps --> MS1125fC1.eps
    }
  \hfill
  \parbox[b]{55mm}{
  \caption{Binary tree structure containing 32 Super-stars (squares). 
    12 Super-stars have been evolved, leaving empty nodes (hexagons).
    In this diagram, the horizontal position of a node reflects the
    radius of the associated super-star. The logical structure of this
    tree is stored in the arrays whose content is displayed on the
    right.}
  \label{fig:bin_tree1}
  }
\end{figure*}

\begin{figure*}
  \centering
  \resizebox{0.93\hsize}{!}{
    \includegraphics{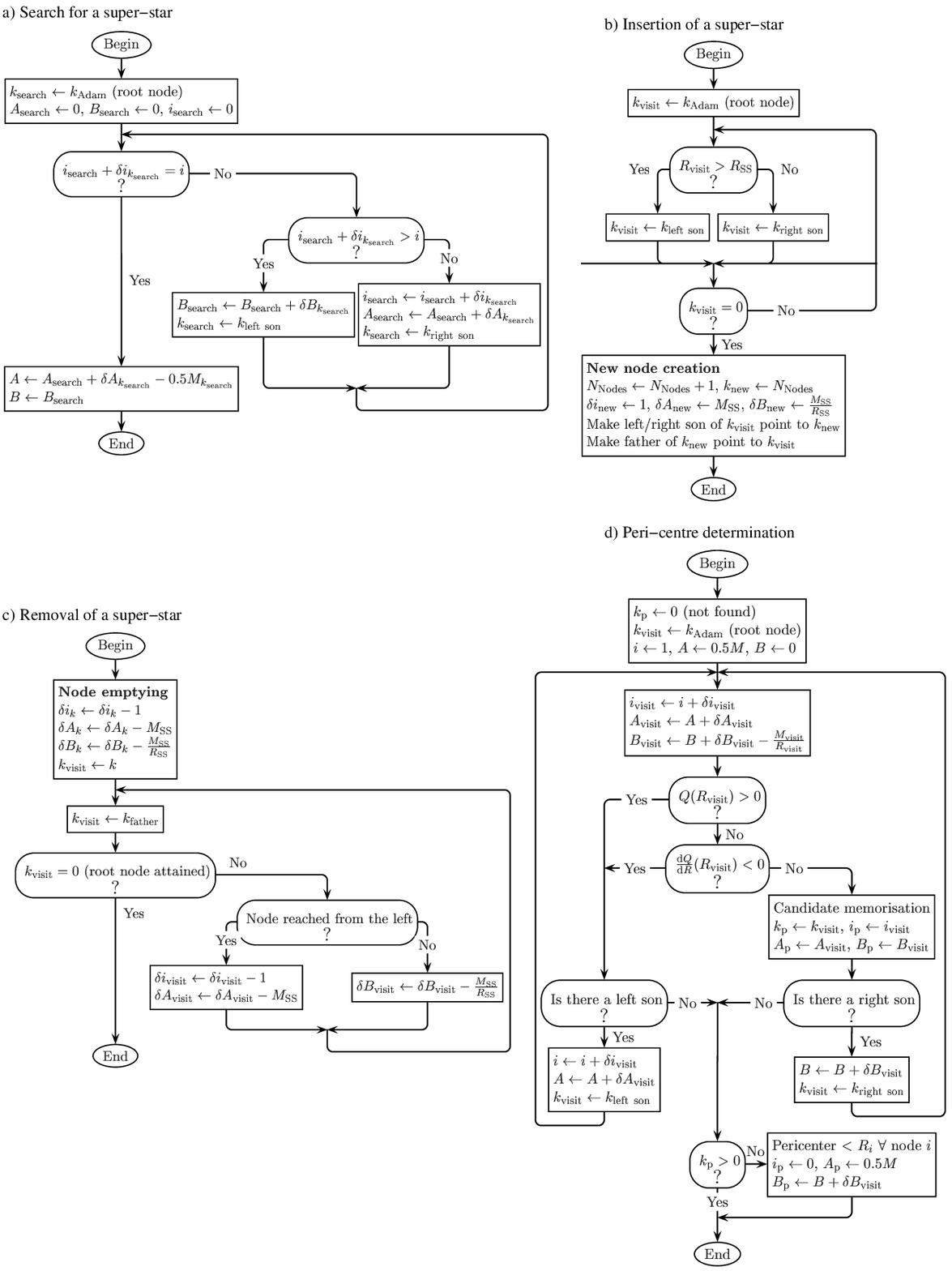}% flows.eps --> MS1125fC2.eps
    }
  \caption{ Flow charts for elementary binary tree routines. {\bf a)} 
    Search for super-star with rank $i$; on exit $A$ and $B$ are its
    potential coefficients (see Sect.~\ref{sec:pot_rep}). {\bf b)}
    Insertion of a super-star of mass $M_{\mathrm{SS}}$ at radius
    $R_{\mathrm{SS}}$. {\bf c)} Extraction of the super-star attached
    to node $k$. {\bf d)} Peri-centre determination for a super-star of mass
    $M$. See text for the definition of $Q(R)$. On exit,
    $k_{\mathrm{p}}$ is the number of the last node with radius
    smaller than the peri-centre radius (in order of increasing
    radius), $i_{\mathrm{p}}$ the corresponding rank and
    $A_{\mathrm{p}}$, $B_{\mathrm{p}}$, the potential coefficients at
    peri-centre.}
  \label{fig:tree_flow}
\end{figure*}
 
In this appendix, we present in some detail the implementation of
the binary tree in charge of storing the potential and ranking
information of the super-star cluster.

The logical structure of the tree is implemented by three integer
arrays: \verb|l_son|, \verb|r_son| and \verb|father|.  \verb|l_son(k)|
is the number of the ``left son'' node of node \verb|k| and so on, see
Fig.~\ref{fig:bin_tree1}.  When the tree is (re-)built, each node
\verb|k| is attributed a super-star \verb|super_star(k)|. When this
particle is evolved and moved to another radius, the node becomes
empty and another leaf node is added to host the super-star. Leaving
empty nodes simplifies the tree update procedures with the cost of
increasing its size. This introduces some numerical overhead as it
causes a faster increase of the number of hierarchical levels
(particularly in the central regions where the time steps are shorter,
see Fig.~\ref{fig:bin_tree1}) but this inconvenience is probably not a
serious concern for the tree is rebuilt from scratch from time to time
in order to keep it reasonably well balanced. This operation is
computationally cheap compared to the numerous tree traversals and
would be called for even if empty nodes were not
created\footnote{However, given that most CPU time is spent in tree
  traversals and that node access probabilities (proportional to
  $\delta t^{-1}$) are highly larger in the centre than in the outer
  parts ($\delta t_\mathrm{max}/\delta t_\mathrm{min} \gg 100$), it is
  quite unfortunate that our ``lazy'' updating method keeps
  lengthening the search path to those most active central particles.
  For the same reason
, building the tree as an ``optimal'' binary
  search tree \citep{Knuth73,Wirth76} instead of a balanced one would
  certainly turn out to be a valuable improvement.}.

To find a super-star with rank $i$ and compute the potential at its
position, one traverses the tree from the root to the corresponding
node using the algorithm sketched in Fig.~\ref{fig:tree_flow}a.  At
the end of this tree traversal, $k_\mathrm{search}$ points to the
proper node and the super-star's potential is $\Phi =
-A/R_{k_\mathrm{search}} - B$. A very similar routine is used to
compute the potential $\Phi(R)$ at any arbitrary radius $R$. Flow
charts for addition/removal of a super-star into/from the binary tree
are depicted in Fig.~\ref{fig:tree_flow}b, c.

Another important role of the binary tree is the computation of peri-
and apo-centre radii for a given super-star, an obvious prerequisite
to the radial placing procedure described in Sect.~\ref{sec:selecRorb}.
Note that a high level of precision is called for: unphysical cluster
evolution could expectedly arise if super-stars' orbits suffered from
any systematic bias. For lack of explicit knowledge of $\Phi(R)$,
solving Eq.~\ref{eq:vrad2eq0} is not straightforward. The main
operation, depicted in Fig.~\ref{fig:tree_flow}d, is a tree traversal
in a search for the node $k_p$ whose radius lies immediately below the
peri-centre, i.e. with $Q(R)=v_\mathrm{rad}^2(R)<0$ and
$\mathrm{d}Q/\mathrm{d}R > 0$. This provides us with the local
$A_\mathrm{p}$ and $B_\mathrm{p}$ potential coefficients. Hence,
computing $R_\mathrm{p}$ reduces to the solution of the simple
equation:
\begin{equation}
  v_\mathrm{rad}^2 = 2E + 2\frac{A_\mathrm{p}+0.5M}{R} + 2B_\mathrm{p}   
  + \frac{J^2}{R^2} = 0.
\end{equation}
Needless to say, the computation of the apo-centre radius is very
similar.

\section{Tests for spurious relaxation}
\label{app:spur_rel}

%------------------------------------------------------------
% Figure : Spurious relaxation
%------------------------------------------------------------

\begin{figure*}
  \resizebox{12cm}{!}{
    \includegraphics{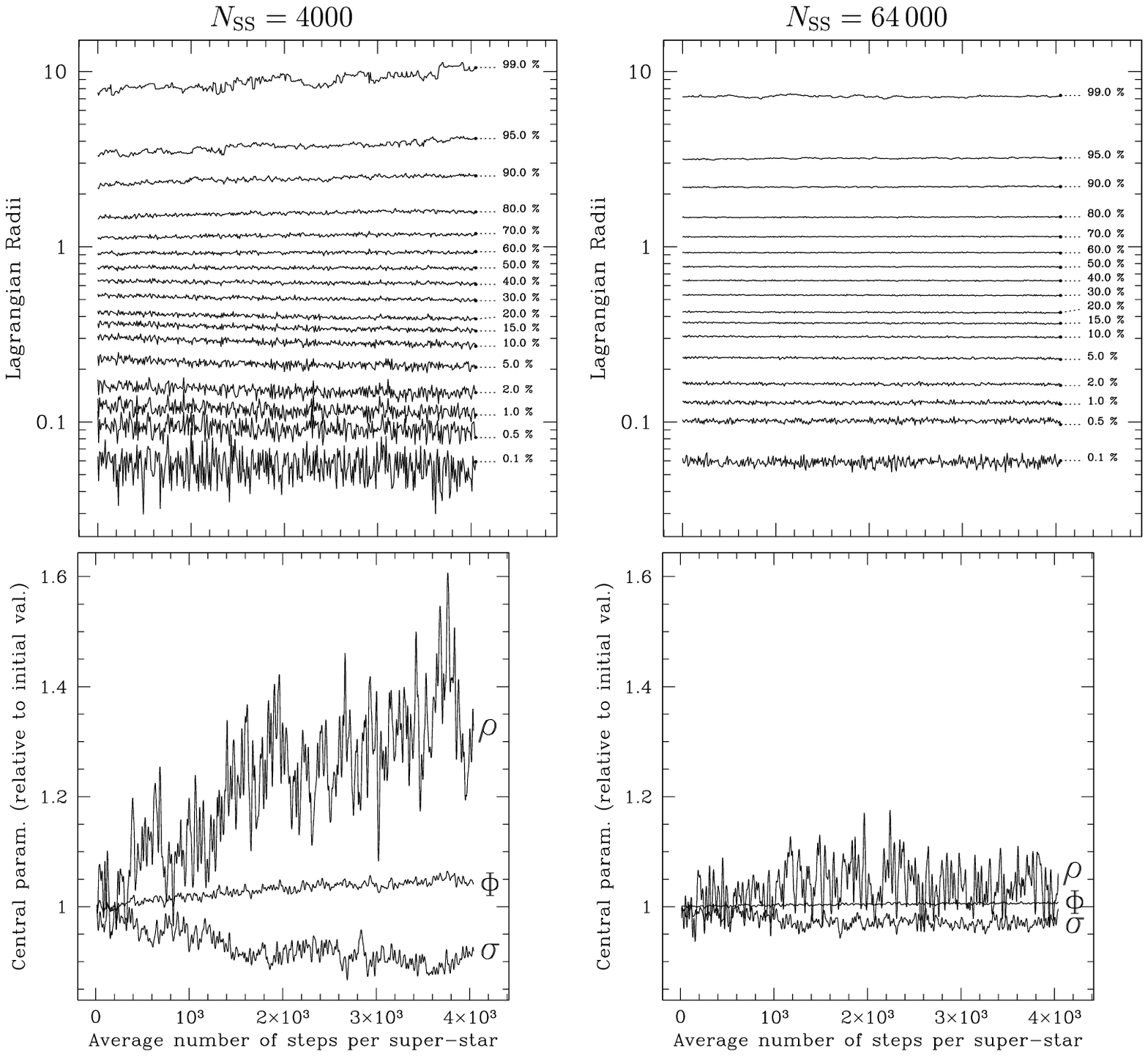}% spurious_rel_nb.eps --> MS1125fD1.eps
    }
  \hfill
  \parbox[b]{55mm}{
  \caption{ Tests for spurious relaxation in Plummer models with 4000 (left) 
    and 64\,000 (right) super-stars. Relaxation was switched off. The
    top panels show the evolution (or absence thereof) of Lagrangian
    radii. The bottom panels present the evolution of central
    properties: density ($\rho$), potential ($\Phi$) and velocity
    dispersion ($\sigma$). These quantities have been normalized by
    the average of the 20 first values of each sequence and the curves
    have been smoothed for the sake of clarity. Note that for
    $N_{\mathrm{SS}}=4000$, some spurious evolution happens while it
    is nearly suppressed for $N_{\mathrm{SS}}=64\,000$.}
  \label{fig:spur_rel}
  }
\end{figure*}

To measure the amplitude of the spurious relaxation due to the fact
that we represent a smooth potential $\Phi$ by a set of super-stars,
i.e. of spherical shells with zero thickness, we carried out a few
simulations in which relaxation was switched off. In such cases, the
algorithm reduces to moving the super-stars on their orbits again and
again. Ideally, the structure of the cluster should not display any
evolution but statistical fluctuations. Figure.~\ref{fig:spur_rel} shows
the results of such tests realized with 4000 and 64\,000 particles.
The computations where stopped when the total number of steps divided
by the number of super-stars reached 4000. For comparison, our
core-collapse simulation for a Plummer model with $2\times 10^6$
particles (Sect.~\ref{subsec:plummer_cc}) required an average of 2300
steps per super-star. From these relaxation-free tests, we conclude
that the amount of spurious relaxation is negligible if the number of
super-stars is of order 16\,000 or larger.

\bibliographystyle{apj}
\bibliography{aamnem99,biblio}

\end{document}